\documentclass[twocolumn]{emulateapj}
\usepackage{epstopdf}
\usepackage{amssymb}
\usepackage{fontenc}
\usepackage{times}
\usepackage{mathptmx}
\usepackage{graphicx}
\usepackage{threeparttable}
\usepackage{longtable}
\usepackage{subfigure}
\usepackage{color}
\usepackage{rotating}

\slugcomment{}

\shorttitle{Neutron-capture Elements in Very-Metal-poor Stars}
\shortauthors{Aoki et al.}

\begin{document}

%% LaTeX will automatically break titles if they run longer than
%% one line. However, you may use \\ to force a line break if
%% you desire.

\title{Diversity of Abundance Patterns of Light Neutron-capture Elements in Very-metal-poor Stars}

%% Use \author, \affil, and the \and command to format
%% author and affiliation information.
%% Note that \email has replaced the old \authoremail command
%% from AASTeX v4.0. You can use \email to mark an email address
%% anywhere in the paper, not just in the front matter.
%% As in the title, use \\ to force line breaks.

\author{Misa Aoki\altaffilmark{1}}
\affil{International Christian University, Mitaka, Tokyo 181-8585, Japan}
\email{g199002a@icu.ac.jp}
\author{Yuhri Ishimaru}
\affil{International Christian University, Mitaka, Tokyo 181-8585, Japan}
\author{Wako Aoki}
\affil{National Astronomical Observatory of Japan, Mitaka, Tokyo 181-8588, Japan}
\author{Shinya Wanajo}
\affil{Department of Engineering and Applied Sciences, Sophia University, Chiyodaku, Tokyo 102-8554, Japan and iTHES Research Group, RIKEN, Wako, Saitama 351-0198, Japan\\
\\
Study based on data collected with the Subaru Telescope, operated by the National Astronomical Observatory of Japan.\\
Received 2016 June 26; revised 2017 January 27; accepted 2017 January 27; published 2017 February 27}

%% Notice that each of these authors has alternate affiliations, which
%% are identified by the \altaffilmark after each name.  Specify alternate
%% affiliation information with \altaffiltext, with one command per each
%% affiliation.

\altaffiltext{1}{}

%% Mark off your abstract in the ``abstract'' environment. In the manuscript
%% style, abstract will output a Received/Accepted line after the
%% title and affiliation information. No date will appear since the author
%% does not have this information. The dates will be filled in by the
%% editorial office after submission. 

\begin{abstract}
We determine the abundances of neutron-capture elements from Sr to Eu for five very-metal-poor stars ($-3$$<$[Fe/H]$<$$-2$) in the Milky Way halo to reveal the origin of light neutron-capture elements. Previous spectroscopic studies have shown evidence of at least two components in the r-process; one referred to as the ``main r-process" and the other as the ``weak r-process," which is mainly responsible for producing heavy and light neutron-capture elements, respectively. Observational studies of metal-poor stars suggest that there is a universal pattern in the main r-process, similar to the abundance pattern of the r-process component of solar-system material. Still, it is uncertain whether the abundance pattern of the weak r-process shows universality or diversity, due to the sparseness of measured light neutron-capture elements. We have detected the key elements, Mo, Ru, and Pd, in five target stars to give an answer to this question. The abundance patterns of light neutron-capture elements from Sr to Pd suggest a diversity in the weak r-process. In particular, scatter in the abundance ratio between Ru and Pd is significant when the abundance patterns are normalized at Zr. Our results are compared with the elemental abundances predicted by nucleosynthesis models of supernovae with parameters such as electron fraction or proto-neutron-star mass, to investigate sources of such diversity in the abundance patterns of light neutron-capture elements. This paper presents that the variation in the abundances of observed stars can be explained with a small range of parameters, which can serve as constraints on future modeling of supernova models.
\end{abstract}

%% Keywords should appear after the \end{abstract} command. The uncommented
%% example has been keyed in ApJ style. See the instructions to authors
%% for the journal to which you are submitting your paper to determine
%% what keyword punctuation is appropriate.


%% From the front matter, we move on to the body of the paper.
%% In the first two sections, notice the use of the natbib \citep
%% and \citet commands to identify citations.  The citations are
%% tied to the reference list via symbolic KEYs. The KEY corresponds
%% to the KEY in the \bibitem in the reference list below. We have
%% chosen the first three characters of the first author's name plus
%% the last two numeral of the year of publication as our KEY for
%% each reference.

%% Authors who wish to have the most important objects in their paper
%% linked in the electronic edition to a data center may do so by tagging
%% their objects with \objectname{} or \object{}.  Each macro takes the
%% object name as its required argument. The optional, square-bracket 
%% argument should be used in cases where the data center identification
%% differs from what is to be printed in the paper.  The text appearing 
%% in curly braces is what will appear in print in the published paper. 
%% If the object name is recognized by the data centers, it will be linked
%% in the electronic edition to the object data available at the data centers  
%%
%% Note that for sources with brackets in their names, e.g. [WEG2004] 14h-090,
%% the brackets must be escaped with backslashes when used in the first
%% square-bracket argument, for instance, \object[\[WEG2004\] 14h-090]{90}).
%%  Otherwise, LaTeX will issue an error. 

\section{Introduction}
One of the major questions in the current nuclear astrophysics is the site of the rapid neutron-capture process (r-process). Very-metal-poor stars are considered to be the key to constrain the site, since their chemical abundances are expected to exhibit the yields of a single to a few nucleosynthetic events \citep[e.g.,][]{mcwill95, ryan96, beers05}. Indeed, a small fraction of very-metal-poor stars show large excesses of neutron-capture elements that are suggested to have been processed by the r-process in the early Galaxy. Previous spectroscopic studies have measured detailed abundance of neutron-capture elements for a considerable number of very-metal-poor stars in the Milky Way halo to identify the origin of the r-process \citep[e.g.][]{honda04, wu15}.

A distinctive feature was revealed from abundance ratios of light and heavy neutron-capture elements of a large sample of metal-poor stars measured by previous studies. Light neutron-capture elements include Sr, Y, and Zr, which locate near the first abundance peak associated with the neutron magic number 50, and those up to Ag in this study. Heavy neutron-capture elements include those near the second abundance peak and heavier elements (e.g., Ba, Eu). The abundance ratios of light and heavy neutron-capture elements such as [Sr/Ba] and [Zr/Ba] show large scatter at low metallicity, suggesting the presence of at least two r-process components \citep[e.g.,][]{burris00, JB02, honda04, aoki05}.  One is well known as the ``main r-process" \citep[e.g.][]{truran02}, which yields both light and heavy neutron-capture elements. A remarkable feature of this process is that it appears to produce a ``universal pattern," being almost identical to the abundance pattern of the solar-system r-process component at least for those heavier than the second abundance peak \citep[e.g.,][]{sneden96, sneden08, barbuy11, siqueira13, siqueira14}. The other is frequently referred to as  the ``weak r-process" \citep[e.g.][]{wana06}, which yields light neutron-capture elements. The latter is not confirmed to be a type of r-process and is sometimes referred to as the Lighter Elements Primary Process \citep[LEPP;][]{trava04}. We note that other processes like the s-process in rapidly rotating massive stars \citep[e.g.,][]{chiap11, cesc13} are also suggested as sources of the scatter in [Sr/Ba]. In this paper, however, we refer to it as the weak r-process according to theoretical predictions for such a process from the studies of supernova nucleosynthesis \citep{wana11, wana13}. 

Previous spectroscopic studies have determined detailed neutron-capture abundances for several metal-poor stars that have excesses of light-to-heavy ratios (e.g., [Sr/Ba]). One of such stars, HD 122563, is a well-studied object that shows deviation of the abundances of neutron-capture elements from the solar r-process abundance distribution \citep[e.g.,][]{honda06}. The abundance pattern of this star has frequently been referred to as a template for the weak r-process. The understanding of the weak r-process abundances is yet insufficient, partially due to the lack of observational constraints. The key elements for the understanding of the weak r-process are those between the first and second abundance peaks, such as Mo, Ru, Pd, and Ag. However, these elements have only weak absorption in the near-UV spectral range, which makes them difficult to measure. Abundances of such elements have been studied for r-process-enhanced stars with very low metallicity \citep[e.g.,][]{ hill02, sneden03}. Pd and Ag are more generally studied for metal-poor stars \citep[e.g.,][]{JB02, hans12}. \citet{peter11} and \citet{hans14} determined abundances of Mo and/or Ru abundances,
%%for a large number of objects
 discussing the origin of these elements assuming the abundance pattern of HD 122563 for the weak r-process. Most of their sample, however, are metal-rich stars, or metal-poor stars with moderate contribution from the main r-process. In this study, we select metal-poor objects that are not main r-process rich stars, and have some excess of light-to-heavy element ratios. Further studies of such stars are required to constrain the abundance patterns of the weak r-process.

Still, it is uncertain whether the abundance of the weak r-process shows a universal pattern as found in the main r-process or shows a diversity. To obtain observational constraints, we analyze the abundance patterns of five very-metal-poor stars showing excesses of light-to-heavy ratios of neutron-capture elements. The spectroscopic data of stars were obtained with the 8.2m Subaru Telescope. Following the methods in previous studies \citep[e.g.,][]{sneden96, honda06}, we measure the stellar abundances of neutron-capture elements and compare the results to the solar r-process pattern. We also compare the abundances of light neutron-capture elements among these stars to inspect the variation of the patterns. 

In \S2, we describe our target stars and details of the spectroscopic observation obtained by the Subaru High Dispersion Spectrograph (HDS). \S3, gives details of the chemical analysis of neutron-capture elements. We present the results and discuss the source of the weak r-process from derived abundances in \S4. Finally, we summarize our study in \S5. 

%--------table1----------------------------------------------------------------------------
\begin{table*}[!t]
\begin{center}
\caption{Object Data}
\begin{tabular}{lcccccr}
\hline
\hline
Object Name & R.A. & Decl. & Exp. Time & S/N&Date&$V_{H}$ \\
& (J2000) & (J2000) & (s) & (4100 \AA)&(UT)&(km $s^{-1}$) \\
\hline
     BD$ +6^{\circ}648$ & 04:13:12.771 & $+$06:35:49.43 & 1800& 150&2007 February 08&-143.5$\pm$0.6\\
    HD 23798 &03:46:45.544 & $-$30:51:25.48 &15,600 & 480&2007 February 08&$89.3\pm$0.6\\
    HD 85773 & 09:53:39.708 & $-$22:50:18.36 & 10,800 & 220&2007 February 08&$147.7\pm0.8$\\
    HD 107752 & 12:22:53.449 & $+$11:36:18.85 & 10,800 & 340&2007 February 08&$219.8\pm0.6$\\
    HD 110184 & 12:40:14.375 & $+$08:31:25.76 &20,400 & 630&2007 February 08&$139.5\pm0.6$\\
    
\hline 
 \end{tabular}
\label{tab:obj}
\end{center}
\end{table*}
%-----------------------------------------------------------------------------------------------------------------------------
\section{Observations}\label{obs}
%--figure1-------------------------------------------------------------------------------------
\begin{figure}[htbp]
\centering
\epsscale{1.0} 
\plotone{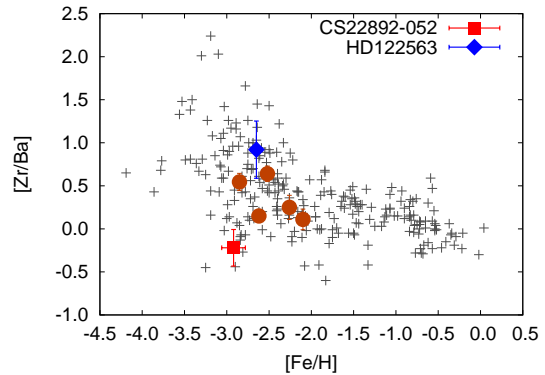}
\caption{[Zr/Ba] abundance ratios as a function of [Fe/H]. The black crosses show data of halo stars taken from the SAGA database (Suda et al. 2008). Red square: CS22892-052; blue diamond: HD~122563; large orange circles: the targets of the present work. The Fe, Z,r and Ba abundances of our target stars are taken from the results of this study.}
\label{zrba}
\end{figure}
%-------------------------------------------------------------------------------------------------------------------------- 

Five bright stars (HD 107752, HD 110184, HD 85773, HD 23798, and BD$ +6^{\circ}648$) are selected for our study based on high-resolution spectra of the UV-blue range. Previous studies show that these objects are very-metal-poor ($-3<$[Fe/H]\footnote{The abundance ratio of two elements (A and B) are defined as [A/B]$=\mbox{log}\frac{X_{A}}{X_B}-\mbox{log}\frac{X_{A\odot}}{X_B\odot}$. The abundance ratio is normalized with the solar abundance of the same elements}$<-2$) red giants that have abundance ratios of [Zr/Ba] higher than the solar (or stars with universal patterns like CS 22892-052) value , but not as extreme as HD 122563 as shown in Figure \ref{zrba}. 

The targets were observed on 2007 February 8th and 9th, with the 8.2m Subaru Telescope HDS \citep{noguchi02}. The wavelength coverage is from 3000 to 4630 {\AA} with a resolving power of $R=$90,000. The signal-to-noise ratio (S/N) of the spectrum is estimated from photon counts at 4100 {\AA}. In addition, we use archived HDS spectra taken from the Japanese Virtual Observatory (JVO)\footnote{http://jvo.nao.ac.jp/portal/top-page.do} for measuring the Ba abundance from lines at longer wavelengths (such as 4554, 4934, 5854, and 6497 {\AA}) that are not covered with our original data. The weak Ba lines at longer wavelengths have little effect of hyperfine splitting and saturation compared to the strong 4554 {\AA} line; therefore, we expect a more accurate measurement from these lines.

Positions of objects, exposure time, S/N, observed dates, and heliocentric radial velocities are summarized in Table \ref{tab:obj}. Radial velocities are measured using Fe lines, and their errors are estimated from standard deviation. The obtained velocities agree well with those obtained by previous studies using high dispersion data. Hence, no signature of radial velocity variation is found from our data. The reduction of the raw data is carried out in a standard process using the IRAF echelle package\footnote{IRAF is distributed by the National Optical Astronomy Observatories, which are operated by the Association of Universities for Research in Astronomy, Inc., with cooperation of the National Science Foundation}. 
%%Honda no S/N????The signal to noise ratio (S/N) of the spectrum (per 0.9 km s−1 pixel) estimated from the photon counts is 140 at 3100; 480 at 3500; 860 at 3900; 1080 at 4200 and 1340 at 4500.

\section{Chemical Abundance Analysis}
Abundance determinations are performed on the basis of the model atmospheres and spectral line data, assuming the local thermodynamic equilibrium (LTE). The model atmospheres with the revised opacity distribution function (NEWODF) by \citet{caskur03} are adopted. We employ a one-dimensional LTE spectral synthesis code for our calculations of synthetic spectra. The code is based on the same assumptions as the model atmosphere program of \citet{tsuji78}, which have been used in previous studies \citep[e.g.,][]{aoki09}. Spectral line data for line identifications in our analysis are adopted from previous studies of \citet{ivans06} and \citet{honda06}. The line list used in the analysis is given in Table \ref{lines}. 

Among the model atmosphere parameters, we adopt effective temperatures ($T_{\rm eff}$) from studies given in Table \ref{tab:stepa}. The exception is HD 107752 for which very different $T_{\rm eff}$ (about 400 K) has been obtained by different studies. We estimate $T_{\rm eff}$ from the color ($V- K$), adopting the Two Micron All Sky Survey $K$ magnitude and $V$ magnitude by \citet{oja85} taken from the SIMBAD database.\footnote{SIMBAD Astronomical Database: http://simbad.u-strasbg.fr/simbad/}
We prefer $V-K$ to $B-V$ for the estimate of $T_{\rm eff}$, because the $V-K$ color of giant stars is less dependent on metallicity and molecular absorption. From the empirical calibration between $T_{\rm eff}$ and $V-K$ color for giant stars by \citet{gonz09}, we estimate the $T_{\rm eff}$ of HD 107752 to be 4760 K which agrees well with the $T_{\rm eff}$ estimated from $B-V$ by \citet{ishig13}. Surface gravity (log {\it g}), metallicity ([Fe/H]) and micro-turbulence ($\xi$) are determined from the analysis of \ion{Fe}{1} and \ion {Fe}{2} lines as their abundances measured using individual lines do not show any trend or systematic differences. The parameters determined for each target are summarized in Table \ref{tab:stepa}. 

%----table2-----------------------------------------------------------------------------------------------
\begin{table*}[p]
\begin{center}
\caption{Line Data and Equivalent Widths}
\begin{tabular}{ccccccp{0.1pt}ccp{0.1pt}ccp{0.1pt}ccp{0.1pt}cc}
\hline
\hline
& & & &\multicolumn{2}{c}{ BD$ +6^{\circ}648$} & & \multicolumn{2}{c}{HD 23798} & & \multicolumn{2}{c}{HD 85773}& & \multicolumn{2}{c}{HD 107752}& & \multicolumn{2}{c}{HD 110184} \\
\cline{5-6}\cline{8-9}\cline{11-12}\cline{14-15}\cline{17-18}   \vspace{-0.2cm}\\
Element& Wavelength& L.E.P& log {\it gf}& log $\epsilon$&EW& &log $\epsilon$ & EW& & log $\epsilon$&EW& & log $\epsilon$& EW& &log $\epsilon$& EW\\
&(\AA) \vspace{0cm} &(eV) &  &  & (m\AA)& & &(m\AA) & & & (m\AA)& & & (m\AA)& & & (m\AA)\\
\hline 
Sr {\tiny II}&4077.71& 0.00 & 0.17 
& \nodata & \nodata &
& \nodata & \nodata &
&$ -0.02$ & 201.4 &
& $-0.13$ & 144.0 &
& \nodata & \nodata \\

Sr {\tiny II}&4215.52& 0.00 & $-0.17$
& 0.95 & 273.2 &
& 0.86 & 266.6 &
& 0.02 & 174.1&
& $-0.39$ & 117.5 &
& 0.46 & 247.4\\		

Y {\tiny II}& 3549.01& 0.13 & $-0.28$
& 0.02 & 85.3 &
&$-0.26$ &83.9&
& $-0.98$  & 63.8 &
&$-0.90$ & 42.6 &
& $-0.91$ &80.6 \\

Y {\tiny II}& 3600.74& 0.18 & 0.28
& \nodata & \nodata &
& \nodata & \nodata &
& \nodata & \nodata &
&\nodata & \nodata &
& $-0.99$ & 92.6\\

Y {\tiny II}& 3611.04 & 0.13 & 0.10
& $-0.16$ & 90.2&
& $-0.07$ & 103.7&
& $-1.17$ & 68.9 &
& $-0.98$ & 52.6 &
& $-0.85$ & 93.3\\

Y {\tiny II}& 3950.36& 0.10 & $-0.49$
& 0.21 & 93.7 &
& 0.03 & 96.6&
& $-0.73$ & 70.5 &
& $-0.84$ & 41.4 &
&$-0.53$ & 89.7 \\

Y {\tiny II}& 4398.01& 0.13 & $-1.00$
 & \nodata & \nodata &
& 0.13 & 82.6 &
& \nodata & \nodata &
&$ -0.78$ & 21.8 &
& \nodata & \nodata \\

Zr {\tiny II}&3438.23& 0.09 & 0.42 
& \nodata & \nodata  &
& \nodata & \nodata  &
& \nodata & \nodata  &
& $-0.32$ & 74.8 &
& \nodata & \nodata \\	

Zr {\tiny II}&3457.56& 0.56 & $-0.53$ 
& \nodata & \nodata  &
&0.63 & 67.9 &
& $-0.25$ & 44.2 &
& $-0.12$ & 23.5 &
&$-0.15$ & 64.0\\

Zr {\tiny II}&3479.02& 0.53 & $-0.69$ 
& \nodata  & \nodata  &
& \nodata  & \nodata  &
& \nodata & \nodata &
&$ -0.26$ & 14.8 & 
& \nodata & \nodata \\

Zr {\tiny II}&3536.94 & 0.36 & $-1.31$ 
& \nodata & \nodata &
& 0.54 & 39.8& 
 & \nodata & \nodata &
 & $-0.22$ & 7.2 &
&$-0.31$ & 30.9\\

Zr {\tiny II}&3573.08 & 0.32 & $-1.04$
& 0.62& 54.3 &
& 0.50 & 54.2&
& $-0.36$ & 29.6 &
& $-0.22$ & 13.6&
& $-0.28$ & 43.1 \\

Zr {\tiny II}&3630.02& 0.36 & $-1.11$
& \nodata  & \nodata &
& 0.54 & 55.1 &
& $-0.25$ & 28.8 &
& $-0.28$ & 9.6 &
& $-0.09$ & 46.2 \\

Zr {\tiny II}&4050.33& 0.71 & $-1.00$
& \nodata & \nodata&
& 0.75 & 47.2 &
 & \nodata & \nodata &
 & \nodata & \nodata &
& \nodata & \nodata  \\

Zr {\tiny II}&4071.09& 1.00 & $-1.60$ 
& \nodata & \nodata &
& 0.99 & 18.7 &
 & \nodata & \nodata &
 & \nodata & \nodata &
& \nodata & \nodata \\

Zr {\tiny II}&4208.99& 0.71 &$ -0.46$ 
& \nodata  & \nodata  &
&0.62 & 73.9&
& $-0.05$ & 47.0 &
&$-0.16$ & 21.7&
& 0.17 & 66.8 \\

Zr {\tiny II}&4317.32& 0.71 & $-1.38$ 
& 0.91 & 33.9 &
& 0.85 & 32.8&
& \nodata & \nodata &
 & \nodata & \nodata &
& 0.23 & 24.2 \\

Zr {\tiny II}&4613.92& 0.97 & $-1.52$ 
& \nodata & \nodata &
& 0.98 & 23.2 &
& \nodata & \nodata &
 & \nodata & \nodata &
& \nodata & \nodata \\

Mo {\tiny I}& 3864.10& 0.00 & $-0.01$ 
& 0.03 & 41.9 &
&$ -0.11$ & 30.0& 
& $-0.97$ & 11.6 &
& $-0.90$ & 2.0 &
&$ -0.70$ & 23.0\\

Ru {\tiny I}& 3498.94& 0.00 & 0.31 
&$ -0.33$ & 30.9 &
&$-0.28$ & 29.1&
& $-1.17$ & 12.2 & 
& $-0.96$& 2.5 &
& $-0.91$ & 21.4\\

Ru {\tiny I}& 3728.03& 0.00 & 0.27 
& $-0.28$ & 31.7 &
& $-0.06$ & 38.0 &
& $-0.84$ & 18.9 &
& $-0.95$ & 2.3 &
& $-0.70$ & 23.0\\				

Pd {\tiny I}& 3404.58 & 0.81 & 0.32
& $-0.78$ & 35.9 &
& $-0.74$ & 35.1&
& $-1.30$ & 26.7 &
& $-1.35$ & 4.1 &
& $-1.22$ & 32.1\\

Ba {\tiny II}& 4554.03& 0.00 & 0.14
&0.10&204.3&
& $-0.23$ &196.4&
& $-0.76$ &170.6&
& $-0.99$ &123.7 &
& $-0.91$ & 182.2\\

Ba {\tiny II}& 4934.10& 0.00 & $-0.16$
&0.15&206.0&
& $-0.23$ &197.4&
&$ -0.74$ &174.7& 
& $-0.92$ & 126.3&
&$ -0.91$ & 184.5\\

Ba {\tiny II}& 5853.69& 0.60 & $-0.91$ 
& 0.30 &93.1&
& $-0.06$&84.1&
& $-0.82$ &50.7&
& $-1.30$ & 10.6 &
& $-1.31$ & 30.2 \\

Ba {\tiny II}& 6141.73& 0.70 & $-0.08$ 
& 0.32 &140.5&
& $-0.01$ &135.5&
& $-0.76$ &104.1&
& \nodata & \nodata&
& \nodata & \nodata \\

Ba {\tiny II}& 6496.91& 0.60 & $-0.38$ 
& \nodata & \nodata &
& 0.04 &134.2&
& \nodata & \nodata &
& $-1.13$ & 45.4 &
& $-1.12$ & 92.5 \\
			 
La {\tiny II}& 3794.77& 0.24 & 0.21 
 & \nodata & \nodata &
 & \nodata & \nodata & 
 & \nodata & \nodata &
&$-1.88$ & 11.0 &
 & \nodata & \nodata \\

La {\tiny II}& 3988.52& 0.40 & 0.21 
 & \nodata & \nodata &
 & \nodata & \nodata & 
 & \nodata & \nodata &
&$-1.85$ & 8.7 &
 & \nodata & \nodata \\

La {\tiny II}& 3995.75& 0.17 & $-0.06$ 
 & \nodata & \nodata &
 & \nodata & \nodata & 
 & \nodata & \nodata &
&$-1.82$ & 9.2 &
 & \nodata & \nodata \\

La {\tiny II}& 4086.71& 0.00 & $-0.07$ 
& $-0.64$ & 71.7 &
&$-1.05$ & 58.6& 
 & \nodata & \nodata &
&$-1.79$ & 14.3 &
& $-1.84$ & 40.6\\

La {\tiny II}& 4123.22& 0.32 & 0.13 
& $-0.70$ & 60.2 &
& $-0.97$ & 55.1 &
& $-1.78$ & 25.6 &
& $-1.80$ & 9.8 &
& $-1.76$ & 33.3\\

La {\tiny II}& 4322.51& 0.17 & $-0.93$ 
& $-0.81$ & 17.2 &
& $-0.94$ & 14.6 &
& \nodata & \nodata &
& \nodata & \nodata &
& $-1.67$ & 8.9 \\

La {\tiny II}& 4429.91& 0.23 & $-0.35$ 
& \nodata & \nodata &
& \nodata & \nodata &
& $-1.64$ & 18.0 &
& \nodata & \nodata &
& $-1.65$ & 23.4\\
			
Ce {\tiny II}& 4014.90& 0.53 & $-0.20$ 
& \nodata & \nodata &
& \nodata & \nodata &
& $-1.33$ & 5.3 &
& \nodata & \nodata &
& \nodata & \nodata \\

Ce {\tiny II}& 4053.50& 0.00 &$-0.61$ 
& \nodata & \nodata &
& \nodata & \nodata &
& \nodata & \nodata &
& $-1.23$ & 3.1 &
& \nodata & \nodata \\

Ce {\tiny II}& 4073.47& 0.48 & 0.21 
& \nodata & \nodata &
& $-0.72$ & 22.0 &
& $-1.52$ & 9.2 &
& \nodata & \nodata &
& \nodata & \nodata \\

Ce {\tiny II}& 4115.37& 0.92 & 0.10 
& \nodata & \nodata &
& \nodata & \nodata &
 & \nodata & \nodata &
&$-1.15$ & 1.6 &
& \nodata & \nodata \\

Ce {\tiny II}& 4165.60& 0.91 & 0.52 
& \nodata & \nodata &
& \nodata & \nodata &
& \nodata & \nodata &
&$ -1.45$ & 2.1 &
& \nodata & \nodata \\

Ce {\tiny II}& 4222.60& 0.12 & $-0.15 $
& $-0.50$ & 35.3 &
& $-0.69$& 28.3&
& $-1.46$ & 12.7 &
&$-1.40$& 4.4 &
& \nodata & \nodata \\

Ce {\tiny II}& 4418.78& 0.86 & 0.27 
& \nodata & \nodata &
&$ -0.67$ & 11.3 &
& \nodata & \nodata &
& \nodata & \nodata &
& $-1.34$& 7.1 \\

Ce {\tiny II}& 4483.89& 0.86 & 0.10
& \nodata & \nodata &
& \nodata & \nodata &
& \nodata & \nodata &
& \nodata & \nodata &
& $-1.14$ & 7.6 \\

Ce {\tiny II}& 4523.08& 0.52 & $-0.08 $
& \nodata & \nodata &
& \nodata & \nodata &
& \nodata & \nodata &
& $-0.99$ & 4.6 &
& \nodata & \nodata \\

Ce {\tiny II}& 4539.74& 0.33 &$ -0.08 $
&$ -0.37$ & 33.4 &
& $-0.62$ & 24.0& 
& \nodata & \nodata &
& $-1.30$ & 3.9 &
& \nodata & \nodata \\

Ce {\tiny II}& 4551.29& 0.74 & $-0.42$ 
& \nodata & \nodata &
& \nodata & \nodata &
& \nodata & \nodata &
& \nodata & \nodata &
& $-1.19$ & 3.3 \\

Ce {\tiny II}& 4562.36& 0.48 & 0.21 
& $-0.36$ & 39.2 &
&$ -0.58$ & 29.8 &
& $-1.30$ & 14.1 &
& $-1.31$ & 4.9 &
& $-1.33$ & 17.6 \\

Ce {\tiny II}& 4628.16& 0.52 & 0.14 
& $-0.36$ & 33.5 &
& \nodata & \nodata &
 & \nodata & \nodata &
& \nodata & \nodata &
& $-1.30$ & 14.9 \\	

Nd {\tiny II}& 4012.70 & 0.00 & $-0.60$ 
& \nodata & \nodata &
& \nodata & \nodata &
 & \nodata & \nodata &
& $-1.20$ & 5.6 &
& \nodata & \nodata \\

Nd {\tiny II}& 4021.33& 0.32 & $-0.10$ 
& $-0.55$ & 14.1 &
& $-0.77$ & 25.0 &
& $-1.43$ & 14.1 &
&$ -1.43$ & 4.4 &
& \nodata & \nodata \\

Nd {\tiny II}& 4059.95& 0.20 & $-0.52$ 
& \nodata  & \nodata  &
&$ -0.76$ & 15.9 &
 & \nodata & \nodata &
& \nodata & \nodata &
& \nodata & \nodata \\

Nd {\tiny II}& 4061.08& 0.47 & 0.55 
&$ -0.35$ & 33.1 &
&$-0.68$ & 53.8 &
& $-1.35$ & 33.1 &
& $-1.31$ & 14.8 &
& \nodata & \nodata  \\

Nd {\tiny II}& 4368.63& 0.06 & $-0.81$ 
& \nodata  & \nodata  &
& $-0.64$ & 18.0 &
&$ -1.24$ & 10.3 &
& \nodata & \nodata &
& $-1.33$ & 11.4 \\

Nd {\tiny II}& 4446.38& 0.20 & $-0.35$ 
&$ -0.42$ & 15.5 &
& $-0.63$ & 29.1 &
& $-1.28$ & 15.9 &
& \nodata & \nodata &
& $-1.32$ & 19.3\\

Nd {\tiny II}& 4501.81 & 0.20 & $-0.69$ 
&\nodata & \nodata &
& \nodata  & \nodata  &
 & \nodata & \nodata &
& $-1.11$ & 3.6 &
& \nodata & \nodata \\

Nd {\tiny II}& 4513.33 & 0.06 & $-1.33$ 
&$ -0.29$ & 5.3 &
& $-0.54$ & 7.5 &
&$ -1.03$ & 5.3 &
& \nodata & \nodata &
& $-1.24$ & 4.8\\

Nd {\tiny II}& 4563.22& 0.18 & $-0.88$ 
&$ -0.27$ & 9.6 &
& \nodata  & \nodata  &
 & \nodata & \nodata &
& $-1.11$ & 2.6 &
& \nodata & \nodata \\

Eu {\tiny II}& 3819.64& 0.00 & 0.51 
& \nodata  & \nodata  &
& $-1.62$ &113.8 &
& $-2.23$ & 78.3 &
& \nodata & \nodata &
& $-2.10$ & 104.7 \\

Eu {\tiny II}& 4129.70 & 0.00 & 0.22 
& $-1.38$ &111.4 &
& $-1.42$ & 110.7 &
& $-2.00$ & 76.1 &
& $-2.01$ & 31.2 &
& -$1.96$ & 94.4 \\

Eu {\tiny II}& 4205.04& 0.00 & 0.21 
& \nodata &\nodata &
& $-1.46$& 125.5 &
& $-2.02$ & 81.2 &
& \nodata & \nodata &
& \nodata & \nodata \\

Pb {\tiny I}& 3683.46& 0.97 & $-0.54$  
&$-0.05$ &7.6&
&$ -0.43$ & 2.7 &
& $<-0.65$ & \nodata &
& \nodata & \nodata &
& \nodata & \nodata \\
\hline
\label{lines}
\end{tabular}
\end{center}
\end{table*}
%-----------------------------------------------------------------------------------
For elements for which several absorption lines are available in our spectra, the averaged abundances measured from individual lines are taken as the final results. On the other hand, for those that have only one or a few measurable lines, the spectrum synthesis technique is applied for the measurement of their abundances. The effects of hyperfine splitting are taken into account in the analysis of Ba, La, and Eu lines. A total of 11 neutron-capture elements from Sr to Eu are measured for each star. We measure the abundances of Mo, Ru, and Pd for the first time for the five stars in our sample. Sr is also obtained for HD 85773, HD 23798, and HD 107752 for the first time. 

%----table3----------------------------------------------------------------------------------------
\begin{table*}[!t]
\begin{center}
\caption{Stellar Parameter}
\begin{threeparttable}
\begin{tabular}{lccccccc}
\hline
\hline
Object Name & $T_{\rm eff}$ & [Fe/H] & log {\it g}\tnote{1} & log {\it g{\scriptsize \_r}} \tnote{2} & $\xi$ & $\xi${\scriptsize \_Ba} \tnote{3} & References \\
& (K) & (dex) & (dex) & (dex) & (km $s^{-1}$) &(km $s^{-1}$)&\\
\hline
    BD$ +6^{\circ}648$ &4400 & $-$2.11 & 0.90 & 1.30 & 1.85 &  1.80 &1 \\
    HD 23798 & 4450 & $-$2.26 & 1.06 & 1.26 & 2.17 & 1.92& 2\\
    HD 85773  & 4268 & $-$2.62 & 0.87 & 0.87 & 1.98 & 2.00& 3\\
    HD107752 & 4760 & $-$2.85 & \nodata & 1.40 & 1.90 & 1.85& \nodata \\
    HD110184 & 4240 & $-$2.52 & 0.30 & 0.50 & 2.21 & 2.32& 4\\
\hline
 \end{tabular}
\begin{tablenotes}\footnotesize
\item[1] log {\it g} from precedent research
\item[2] log {\it g} regulated from abundances derived from Fe I and Fe II
\item[3] $\xi$ determined from JVO data used for Ba abundance measurements
\tablerefs{(1)\citet{aoki08}, (2) Burris et al. (2000), (3) Simmerer et al. (2004),  (4) Honda et al. 2004}
\end{tablenotes}
\end{threeparttable}
\label{tab:stepa}
\end{center}
\end{table*}
%-----------------------------------------------------------------------------------------------------------------------------
The Pb abundance is sensitive to the contributions of the s-process, in particular, at low metallicity. In order to verify the contribution of the s-process to the targets, the Pb 3683 and 4058 {\AA} lines are inspected for our objects. 

In the following subsections, details of the analysis and results are described. The abundances determined from individual lines are given in Table \ref{lines}. Abundance results are given in Table \ref{tab:ElAbund}. To derive the [X/Fe] values, we use the solar-system abundances obtained by \citet{asp09}.
%-------figure2--------------------------------------------------------------------------------------
\begin{figure*}[!b]
\centering
\subfigure{
        \resizebox{90mm}{!}{\includegraphics{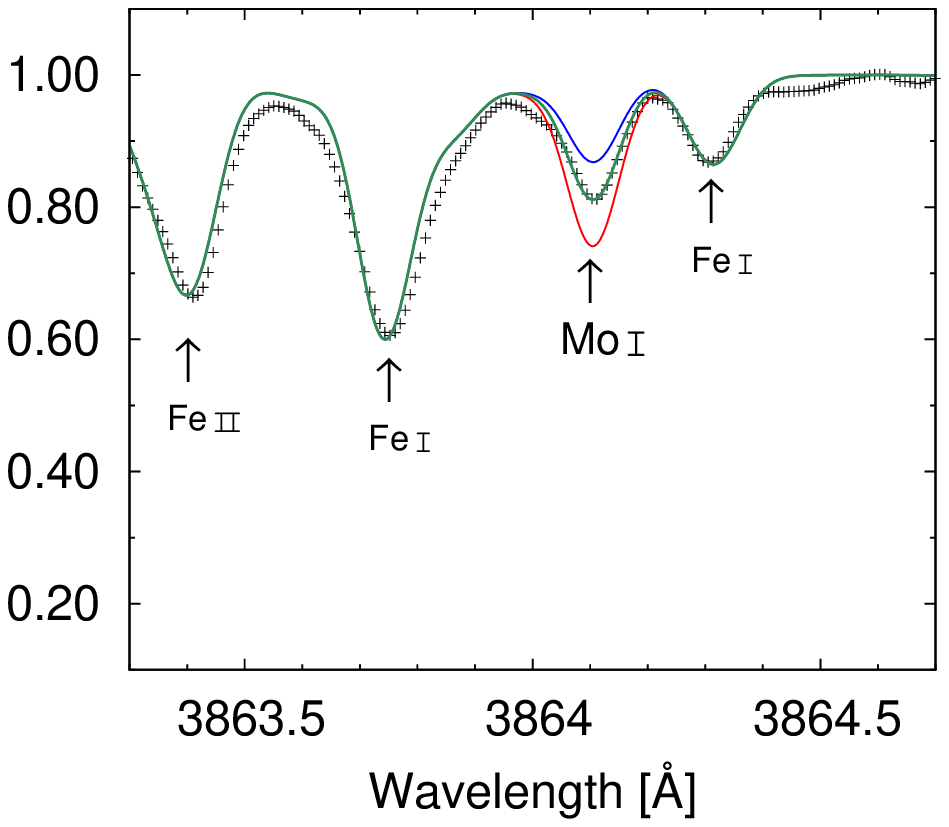}}
\hspace{-17mm}
        \resizebox{90mm}{!}{\includegraphics{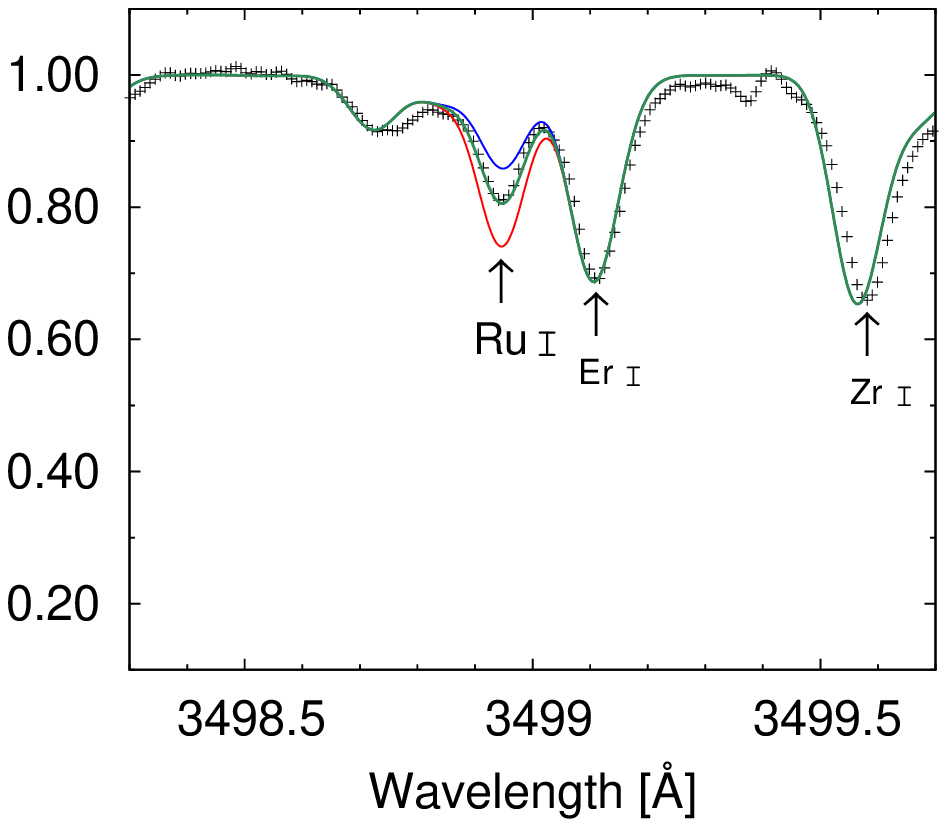}}
}
\subfigure{
        \resizebox{90mm}{!}{\includegraphics{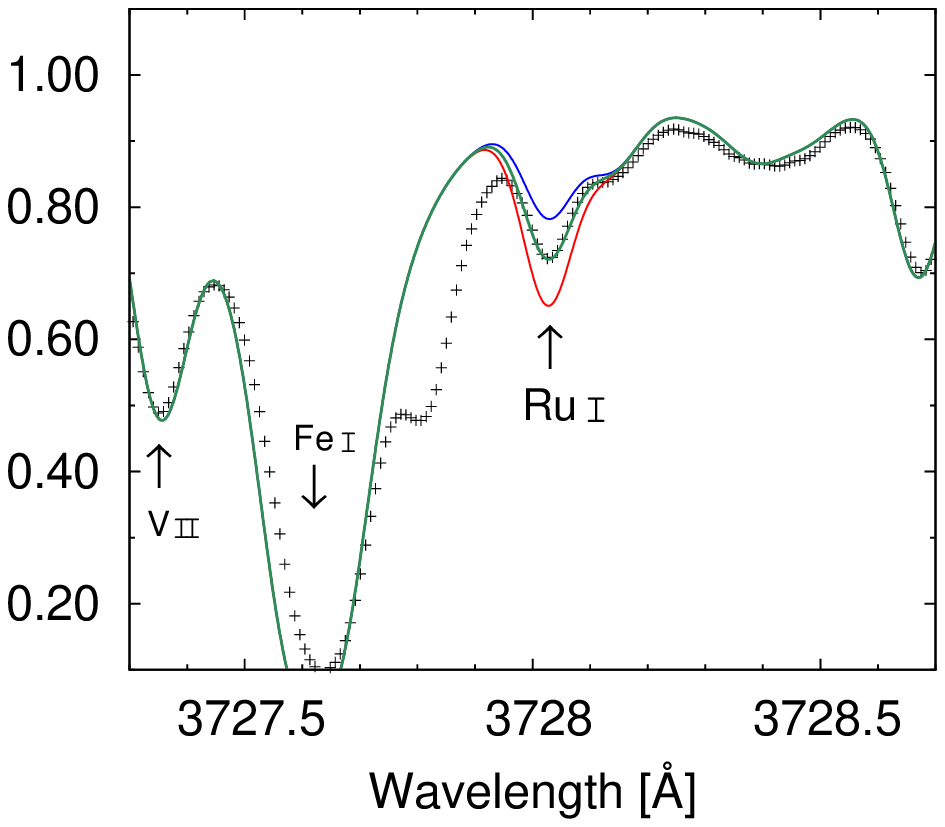}}
\hspace{-17mm}
        \resizebox{90mm}{!}{\includegraphics{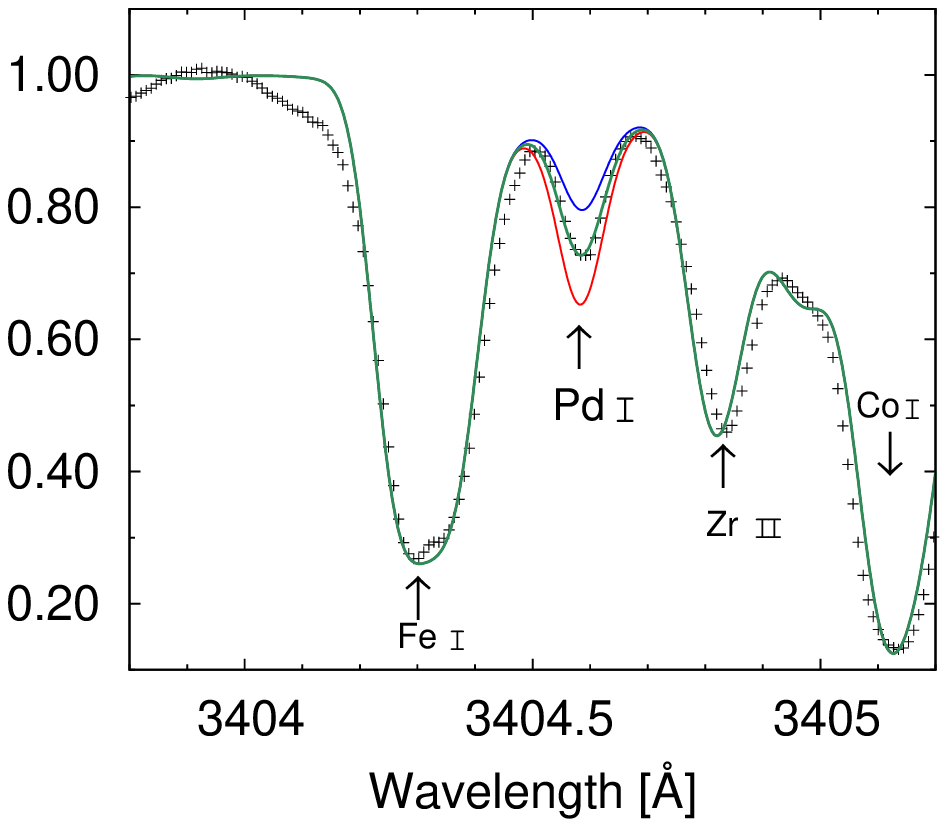}}
}
\caption{Observed light neutron-capture elements of HD 110184: the \ion{Mo}{1} line (3864.10 {\AA}), two \ion{Ru}{1} lines (3498.94 and 3728.03 {\AA}) and the \ion{Pd}{1} line (3404.58 {\AA}). Dotted lines are the observations; solid green (middle) lines are the spectra calculated for the adopted abundances. The blue (upper) and red (lower) lines are the spectra calculated changing the values by $\pm$0.2 dex.}
\label{fig:lighte11}
\end{figure*}
%-----------------------------------------------------------------------------------------------------------------------------

\subsection{Light Neutron-capture Elements (38$\leq$Z$\leq$48)}
The Sr abundance is measured using two strong \ion{Sr}{2} lines at 4078 and 4216 {\AA} fitting the synthetic spectra to our observed spectra. In the fitting process, line broadening affected by the instrument, macro-turbulence, and rotation is approximated by a Gaussian profile. The Sr abundance determined from strong lines is not as certain as abundances of other elements due to the saturation effect. Hence, we do not include the abundance of this element in later sections, where we discuss the light neutron-capture elements. 

%--table4-3------------------------------------------------
\begin{sidewaystable*}[p]
\caption{Elemental Abundances}
\begin{tabular}{lrrrrrp{0.01pt}rrrrrp{0.01pt}rrrrrp{0.01pt}rrrrrp{0.01pt}rrrrr}
\hline
\hline
& \multicolumn{5}{c}{BD$ +6^{\circ}648$}&& \multicolumn{5}{c}{HD 23798} && \multicolumn{5}{c}{HD 85773}&& \multicolumn{5}{c}{HD 107752}&& \multicolumn{4}{c}{HD 110184} \\ 
\cline{2-6} \cline{8-12} \cline{14-18} \cline{20-24} \cline{26-30} \\
Species & log$\epsilon$ & $\sigma_{ran}$&$\sigma_{tot}$ &[X/Fe] &{\it N} &&
log $\epsilon$ &$\sigma_{ran}$ &$\sigma_{tot}$ &[X/Fe]  &{\it N} && 
log $\epsilon$&$\sigma_{ran}$ &$\sigma_{tot}$ &[X/Fe] & {\it N}&&
log $\epsilon$ & $\sigma_{ran}$ &$\sigma_{tot}$ &[X/Fe] &{\it N} &&
log $\epsilon$ & $\sigma_{ran}$ &$\sigma_{tot}$ &[X/Fe] & {\it N} \\
\hline
Sr {\tiny II} & 
0.95 & 0.15 &0.19 &0.19& 2 & &  
0.86 &0.15&0.19&0.25& 2 &&
0.00 & 0.16 &0.20&$ -0.25$ & 2 &&  
$-0.26$ & 0.15 &0.28& $-0.28$ & 2 & 
 &0.46 &0.15 &0.28& 0.11 & 1  \\

Y {\tiny II}& 
\,\,0.02 &0.15 &0.26& $-0.09$ & 3 & &  
$-0.04$ & 0.15 &0.26&\,\,0.01 &4& &
$-0.96$ &0.18 &0.28&\,\,$-0.55$ &3 & &  
 $-0.87$ & 0.15 &0.16&  $-0.23$& 4   & &  
$-0.82$ &0.17 &0.18& $-0.51$ & 4  \\

Zr {\tiny II}& 
\,\,0.76 & 0.15 &0.18&\,\,0.29 & 2 & &  
\,\,0.71 & 0.18 &0.20&\,\,0.39 &9 &&
$-0.23$& 0.16&0.19& $-0.19$ & 4 & &  
$-0.22$ & 0.15 &0.16& 0.05 & 7   & &  
$-0.07$ & 0.21&0.21 & $-0.13$ & 6  \\

Mo {\tiny I}& 
\,\,0.03 & 0.15 &0.25&\,\,0.25 & 1 &&  
$-0.11$ & 0.15 &0.25 &\,\,0.27 & 1 &&
$-0.97$ &0.16 &0.26& $-0.23$ & 1& &  
$-0.90$ & 0.15 &0.17& 0.07& 1   & &  
$-0.70$ &0.15 & 0.17& $-0.06$ &1  \\

Ru {\tiny I}& 
$-0.31$ & 0.15 &0.28& \,\,0.05 & 2   & &  
$-0.17$ & 0.15 &0.28&\,\,0.34 & 2 &&
$-1.01$ &0.16&0.28 & $-0.14$ &2 & &  
$-0.96$ &0.15&0.21& 0.14& 2   & &  
$-0.85$ &0.15 & 0.21&$-0.08$ & 2  \\

Pd {\tiny I}& 
$-0.78$ & 0.15&0.21 &$-0.25$ & 1  & &  
$-0.74$ & 0.15&0.21  &$-0.05$ &  1 & &
$-1.30$ &0.16& 0.22&$-0.25$ & 1  &&  
$-1.35$&0.15& 0.18&$-0.07$& 1 & &  
$-1.22$ &0.15&0.18 & $-0.27$ & 1 \\

Ba {\tiny II}& 
\,\,0.22 &0.15 &0.18&\,\,0.15 & 4 & &  
\,\,$-0.10$ &0.15 &0.18&\,\,$-0.02$ & 5 & &
$-0.77$ & 0.16 & 0.18&$-0.33$ & 4  & &  
$-1.09$ & 0.15 &0.16& $-0.52$ & 4   & &  
$-1.06$ &0.17 &0.18 & $-0.72$ & 4  \\

La {\tiny II}& 
$-0.72$ & 0.15 &0.17&\,\,0.28 & 3 & &  
$-0.99$ & 0.15 &0.17&\,\,0.17  & 3& &
$-1.75$ & 0.16 &0.18&$-0.23$ & 3 & &  
$-1.83$ & 0.15 &0.16& $-0.08$ & 2   & &  
$-1.73$ &0.15& 0.16&$-0.31$ & 4  \\

Ce {\tiny II}& 
$-0.40$ & 0.15&0.18 &\,\,0.13 & 4& &  
$-0.66$& 0.15&0.18&\,\,0.02& 5 & &
$-1.40$ & 0.16&0.19& $-0.36$ & 4  & &  
$-1.26$ & 0.15 &0.16& 0.01 & 7  & &  
$-1.26$ &0.15& 0.16& $-0.32$ & 5  \\

Nd {\tiny II}& 
$-0.38$ & 0.15 &0.18&\,\,0.30 & 5 & &  
$-0.67$ & 0.15 &0.18&\,\,0.17 & 6 & &
$-1.27 $&0.16&0.19&$-0.07$ & 5 & &  
$-1.23 $&0.15 &0.16& 0.20 & 5   & &  
$-1.30$ & 0.15&0.16& $-0.20$ & 3  \\

Eu {\tiny II}& 
$-1.38$  &0.15 &0.19&\,\,0.20 & 1 & &  
$-1.50$ & 0.15&0.19&\,\,0.24& 3& &
$-2.08$ & 0.16&0.20&\,\,0.02 & 3  & &  
$-2.02$  & 0.15 &0.16& 0.31& 1   & &  
$-2.03$ & 0.15&0.16& $-0.03$ & 2\\
\hline
\label{tab:ElAbund}
\end{tabular}
\end{sidewaystable*}
%-----------------------------------------------------------------------------------------------------------------------------
Spectrum synthesis technique is applied to determine the Mo, Ru, and Pd abundance. Examples of detected spectral lines in HD 110184 are shown in Figure~ \ref{fig:lighte11}. The Mo abundance is determined using the \ion{Mo}{1} 3864 {\AA} line. The line is almost free from contamination by other lines (Fig. \ref{fig:lighte11}). Two \ion{Ru}{1} lines at 3499 and 3728 {\AA} are detected, and contamination by other lines is also checked. Mo and Ru are detected for all the objects. The Pd abundance is also measured for the target stars, using the 3405 {\AA} line. We checked the blending of other atomic lines using the atomic line list by \citet{kur93}. The spectra are analyzed using the line list around this \ion{Pd}{1} line provided by \citet{JB02}. 

\subsection{Heavy Neutron-capture Elements (56$\leq$Z$\leq$63)}
The abundance of Ba is determined including the effect of hyperfine splitting \citep{mcwill98}. The quality of HDS spectra from JVO for measuring 4934, 5854, and 6497 {\AA} Ba lines is sufficient for our measurement. The isotope ratios of Ba estimated for the r-process component of solar-system material \citep{sneden96} are assumed. The heavy neutron-capture elements including Ba exhibit enhancement in three out of five stars, suggesting the s-process contribution to Ba in these stars, as discussed in \S \ref{Comp-chap}. We also attempt to determine the abundance assuming the s-process isotope ratios \citep{sneden96} to investigate the effect on derived abundance. The abundance derived assuming the s-process isotope ratio is 0.1 dex greater at most than the abundance derived from those for the r-process case. In particular, the effects are negligible in the analysis of weak Ba lines in the red range. Hence, we employ the results obtained assuming the solar-system r-process isotope ratio as the final results. 

We update the line list of \citet{honda06} for Ce, by adopting {\it gf}-values determined by \citet{lawl09}. The difference of {\it gf}-values between the new and previous line lists is around 0.1 dex, and at most 0.34 dex for the 4015 {\AA} line. This line is used for the determination of the Ce abundance of HD 85773; the abundance of Ce of this star decreased by 0.05 dex after adopting the {\it gf}-values by \citet{lawl09}. Overall, the update of {\it gf}-values has little impact on the final results. 

We have detected \ion{Eu}{2} lines at 3820, 4130, and 4205 {\AA}. Hyperfine splitting is taken into account, adopting the line data of \citet{lawl01}. We assume the isotope fraction of $^{151}\mbox{Eu}$ and $^{153}\mbox{Eu}$ to be 0.5, which is close to the solar isotope fraction.

\begin{figure*}[!t]
\subfigure{
        \resizebox{90mm}{!}{\includegraphics{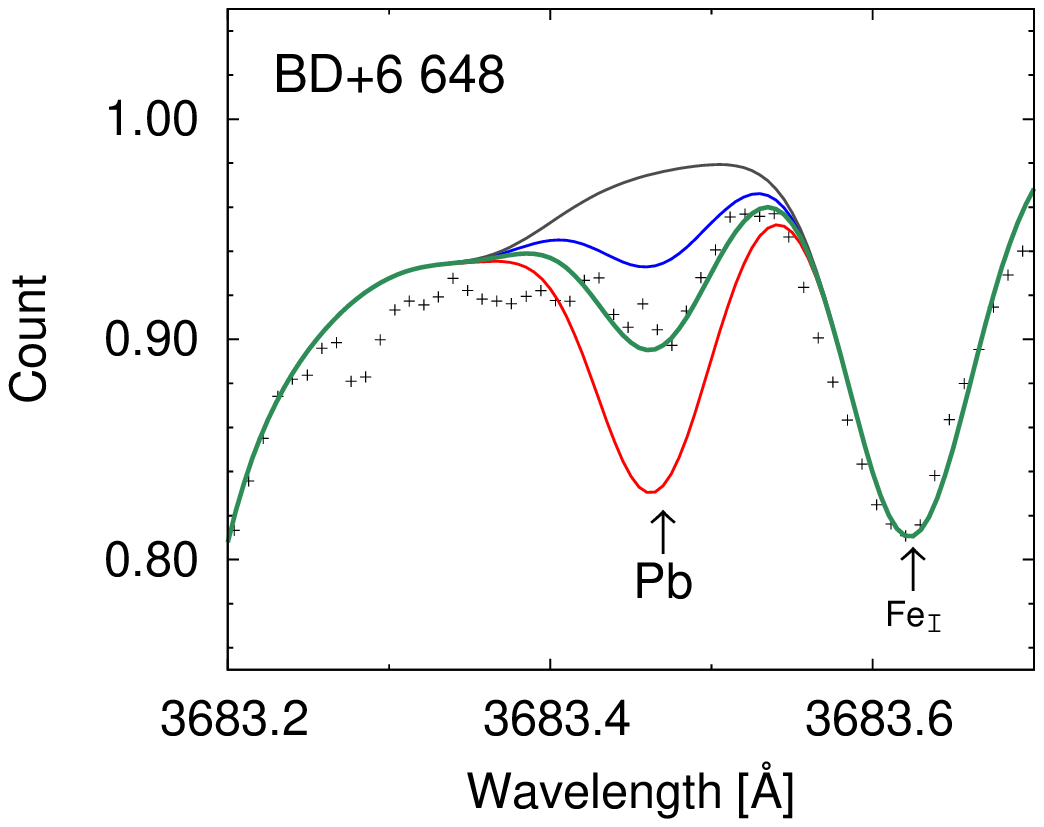}}
\hspace{-1mm}
        \resizebox{90mm}{!}{\includegraphics{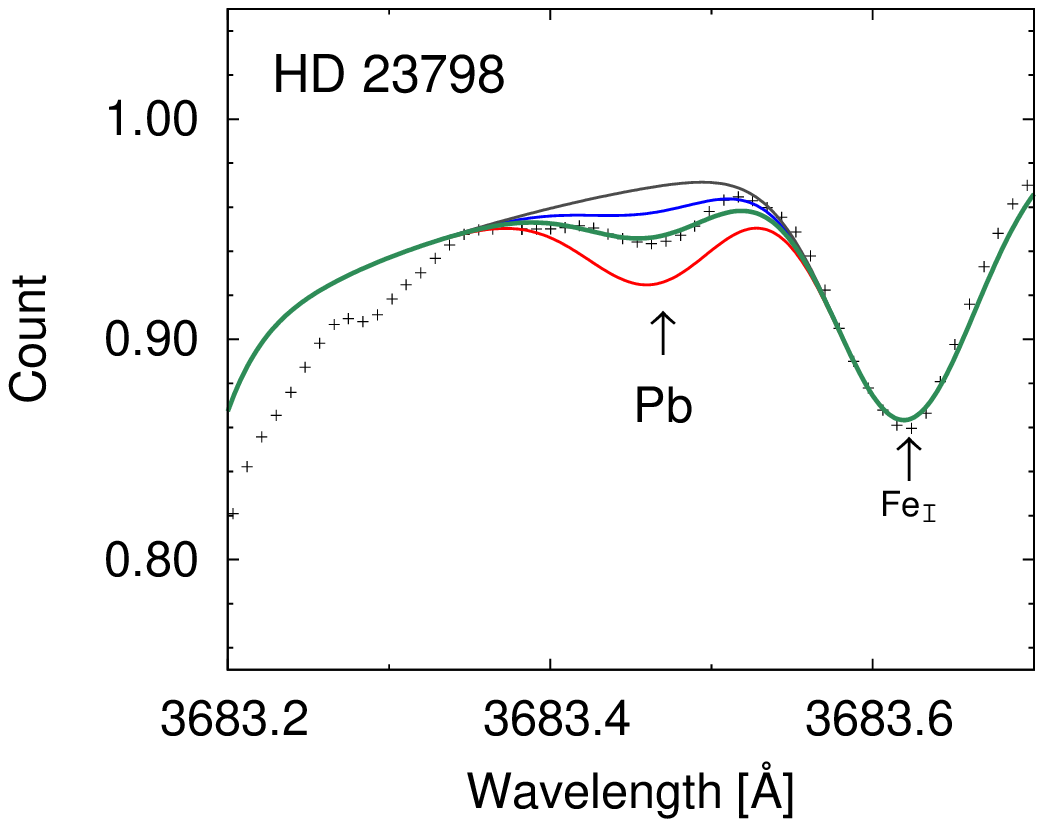}}
}
\subfigure{
        \resizebox{90mm}{!}{\includegraphics{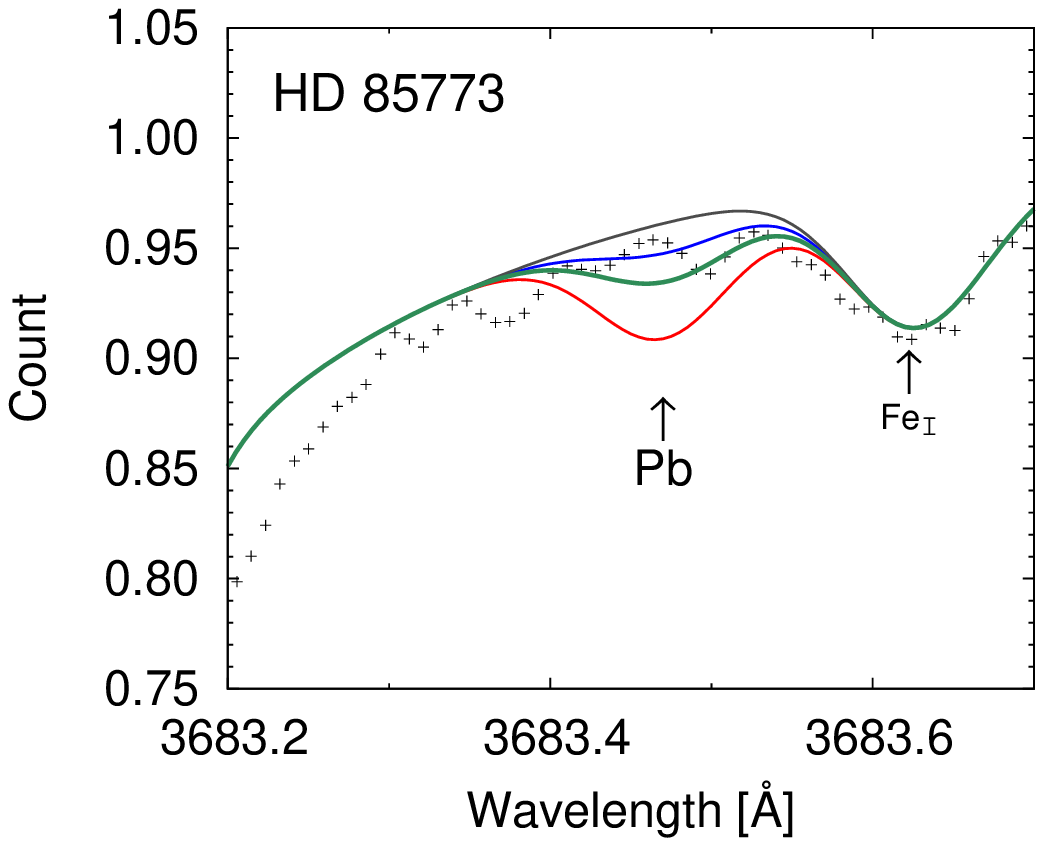}}}
\caption{Observed \ion{Pb}{1} lines of BD$ +6^{\circ}648$, HD 23798 and HD 85773. Dotted line: observations; solid green lines are the spectra calculated for the adopted abundances. The green line is adopted as the upper limit of the Pb abundance for HD 85773. Blue (upper) and red (lower) lines are the spectra calculated changing the values by $\pm$0.3. The gray solid line (top) shows the calculated spectra without Pb contamination. }
\label{pbline}
\end{figure*}

Pb abundance is measured using the \ion{Pb}{1} 3684 {\AA} line. We are able to measure the Pb abundance of two stars, BD$ +6^{\circ}648$ and HD 23798 (Fig.~\ref{pbline}). An upper limit of Pb is estimated for HD 85773 (Fig.~\ref{pbline}). The \ion{Pb}{1} 4058 {\AA} line is not used because of the severe blend with CH molecular lines. The abundance of Pb of BD$ +6^{\circ}648$ was measured using the 4058 {\AA} line in a previous study by \citet{aoki08}. Although the abundance is measured using different lines, our result is in good agreement with their Pb abundance. No meaningful upper limit is obtained for the other two objects. 
%---figure3-------------------------------------------------------------------------------
%---------------------------------------------------------------------------------------------------------------------------

\subsection{Re-analysis of HD 122563}
To verify the consistency of our analysis with previous studies, we apply our analysis procedure to the data of HD122563 reported by \citet{honda06}. We adopt the equivalent widths of \citet{honda06} as well as atomic data, except for Ce. For this element, atomic data updated by \citet{lawl09} are adopted. The results are given in Table \ref{tab:honda} for comparison with those obtained by \citet{honda06}.

The chemical abundances of HD 122563 obtained by our analysis agree well with those obtained by \citet{honda06}. The abundances of most elements are systematically lower by around 0.05--0.10 dex than those obtained by \citet{honda06}. This difference can be attributed to the difference in model atmospheres. \citet{honda06} adopts models provided by \citet{kur93} in which convective overshooting is assumed, while we adopt models of \citet{caskur03}, which are calculated assuming pure mixing-length convection with the new opacity distribution functions. The temperature structure of the models of \citet{caskur03} is several tenths of kelvin lower than those of \citet{kur93}.  

Aside from the effect of the difference in the model atmosphere, Ce abundance shows a relatively large difference (0.19 dex). The updated atomic data by \citet{lawl09} partially explains the discrepancy between the two studies, though the effect is less than 0.1 dex as mentioned above. 

\subsection{Error Estimates}
The values adopted for error bars (total error) for comparisons of abundance patterns are estimated from the random errors and the systematic errors affected by uncertainties of model atmosphere parameters. We estimated the random errors from the standard deviation of abundances derived from individual lines. As for the abundances of elements that have only one or two available lines, the standard deviation of Fe 
%------table5--------------------------------------------------------------------------
\begin{table}[!h]
\centering
\caption{Abundance of HD~122563}
\begin{tabular}{lcp{0.1pt}cp{0.1pt}c}
\hline
\hline
&&&\multicolumn{3}{c}{log $\epsilon$}\\
\cline{4-6}\\
Species& Z & &Our Measurement& &Honda et al. (2006)\\
\hline
Sr {\scriptsize II}& 38&&
$-0.17$ & & $-0.11$ \\

Y {\scriptsize II}& 39&&
$-0.99$ & & $-0.93$ \\

Zr {\scriptsize II}& 40&&
$-0.35$ &  & $-0.28$  \\

Mo {\scriptsize I}& 42&&
$-0.91$ && $-0.87$  \\

Ru {\scriptsize I}& 44&&
$-0.87$ & & $-0.86$ \\
 
Pd {\scriptsize I}& 46&&
$-1.38$ & & $-1.36$ \\

Ba {\scriptsize II}& 56&&
$-1.66$ & & $-1.62$ \\

La {\scriptsize II}& 57&&
$-2.73$& & $-2.66$ \\

Ce {\scriptsize II}& 58&&
$-2.02$ & & $-1.83$ \\

Nd {\scriptsize II}& 60&&
$-2.05$ & & $-2.01$ \\

Eu {\scriptsize II}& 63&&
$-2.84$ &  & $-2.77$  \\
\hline
\label{tab:honda}
\end{tabular}
\end{table}
%----------------------------------------------------------------------------------------
abundance ratios from individual \ion{Fe}{1}  lines is adopted. The standard deviation of Fe is also adopted for elements that have several available lines but the standard deviation of the abundances derived from these lines is smaller than that of Fe. We also independently estimated errors due to the uncertainties of equivalent width measurements based on the relation $\sigma_{w}\simeq(\lambda n^{1/2}_{pix})/(R[S/N])$ \citep{norris01}. For example, the error from the uncertainties of equivalent width of the Mo line at 3864 {\AA} of BD$ +6^{\circ}648$, which has the lowest S/N at 4100 {\AA} among the sample as shown in Table \ref{tab:obj}, is 0.10 dex. The errors due to the uncertainty of continuum placement, which is at most 2\%, is 0.07 dex. These errors are smaller than those estimated from the standard deviation described above. Therefore, we adopted the standard deviations as the random errors, listed as ``$\sigma_{ran}$" in Table \ref{tab:ElAbund}. 
%------table6----------------------------------------------------------------------------------
\begin{table*}[!t]
\begin{center}
\caption{ Error Estimates for HD 107752}
\begin{tabular}{lrrp{1pt}rrp{1pt}rrp{1pt}rr}
\hline
\hline
& \multicolumn{2}{c}{$\Delta T_{\rm eff}$} & & \multicolumn{2}{c}{$\Delta$log {\it g}} & & \multicolumn{2}{c}{$\Delta$[Fe/H]} & & \multicolumn{2}{c}{$\Delta$$\xi$} \\
\cline{2-3} \cline{5-6} \cline{8-9} \cline{11-12}
   Species  & $-100$ K & +100 K & & $-0.3$ & +0.3 & & $-0.3$ & +0.3 & & $-0.3$ & +0.3 \\
\hline
 \ion{Fe}{1}& $-0.12$ & 0.12 & & 0.04 & $-0.04$ & & 0.00 & $-0.01$ & & 0.14 & $-0.12$ \\
\ion{Fe}{2}& $-0.01$ & 0.02 & & $-0.09$ & 0.10 & & 0.00 & 0.01 & & 0.02 & $-0.02$ \\
 \ion{Sr}{2}&$-0.08$ &0.10 & &$-0.03$ & 0.05 & &0.09&$-0.07$& & 0.26 & $-0.22$ \\
\ion{Y}{2}& $-0.08$ & 0.08 & & $-0.07$ & 0.08 & & $-0.01$ & 0.01 & & 0.07 & $-0.05$ \\
 \ion{Zr}{2}&$-0.08$ & 0.08 & & $-0.08$ & 0.09 & & $-0.01$ & 0.01 & & 0.05 & $-0.04$ \\
\ion{Mo}{1}& $-0.15$ & 0.15 & & 0.01 & $-0.01$ & & 0.00 & $-0.02$ & & 0.00 & 0.00 \\
\ion{Ru}{1}& $-0.24$ & 0.18 & & 0.00 & $-0.01$ & & 0.02 & $-0.04$ & & 0.01 & $-0.02$ \\
\ion{Pd}{1}& $-0.17$ & 0.15 & & 0.02 & $-0.03$ & & 0.02 & 0.00 & & 0.02 & $-0.01$ \\
\ion{Ba}{2}& $-0.07$ & 0.11 & & $-0.10$ & 0.09 & & $-0.03$ & 0.04 & & 0.04 & $-0.02$ \\
\ion{La}{2}& $-0.08$ & 0.08 & & $-0.09$ & 0.10 & & $-0.01$ & 0.01 & & 0.00 & $-0.01$ \\
\ion{Ce}{2}& $-0.08$ & 0.08 & & $-0.09$ & 0.09 & & $-0.01$ & 0.01 & & 0.00 & 0.00 \\
\ion{Nd}{2}& $-0.08$ & 0.08 & & $-0.09$ & 0.09 & & $-0.01$ & 0.01 & & 0.01 & 0.00 \\
\ion{Eu}{2}& $-0.08$ & 0.07 & & $-0.09$ & 0.09 & & $-0.01$ & 0.00 & & 0.00& 0.00 \\
\hline 
 \end{tabular}
%% Any table notes must follow the \end{tabular} command.
%%\tablenotetext{a}{Sample footnote for table~\ref{tbl-2} that was generated with the \LaTeX\ table environment}
%%\tablenotetext{b}{Yet another sample footnote for table~\ref{tbl-2}}
%%\tablecomments{We can also attach a long-ish paragraph of explanatory material to a table.}
\label{er10a}
\end{center}
\end{table*}
%---------------------------------------------------------------------------------

The uncertainties of model atmosphere parameters have effects mostly on the systematic errors of the abundances of neutron-capture elements. The 
effects of stellar parameters on abundances are given in Table \ref{er10a}  and Table \ref{er6648a} for HD 107752 ([Fe/H]=$-$3.01) and BD$ +6^{\circ}648$ ([Fe/H]=$-$2.10), respectively, which hold the lowest and highest metallicity in our five samples. The rest of the stars, which hold similar properties, are expected to involve similar errors. Tables show differences in abundance measurements (original abundance $-$ abundance measured using changed parameter) by changing $\pm$100 K for $T_{\rm eff}$, $\pm$0.3 dex for log $g$, $\pm$0.3 dex for [Fe/H], and $\pm$0.3 km s$^{-1}$ for $\xi$.

In particular, effects of changing [Fe/H] and log $g$ are systematically different between abundances derived from neutral and ionized species. Abundances of Mo, Ru, and Pd are determined by lines of neutral species, whereas other heavy elements are measured from ionized species. This indicates that the abundance ratios derived from neutral and ionized species (e.g., Mo/Zr) could be significantly affected by parameters (e.g., $T_{\rm eff}$) adopted in the analysis. We investigate the effect on the abundance pattern of HD 107752, by conducting abundance analyses for $T_{\rm eff}=4400$ and 4760 K. 
%%We determine other parameters for each case. 
The log $g$, [Fe/H], and $\xi$ are derived for each $T_{\rm eff}$, resulting in 0.7dex higher log $g$ for the higher $T_{\rm eff}$. The abundance ratios of heavy elements are  $\sim 0.5$ dex systematically greater for 4760 K. Interestingly, however, the abundance patterns are not significantly dependent on the choice of $T_{\rm eff}$. The effect of the changes of $T_{\rm eff}$ adopted in the analysis is larger for elements measured from neutral species. On the other hand, changes of log $g$ have little effect on the abundances of these elements, while elements measured from ionized species are affected by changes of log $g$ as well by the changes of $T_{\rm eff}$. Hence, even though both neutral and ionized species are studied, the overall abundance pattern obtained by the present work is robust.
We added the random and systematic errors in quadrature to derive the values adopted for error bars for comparison of abundance patterns, which are listed as ``$\sigma_{tot}$" in Table \ref{tab:ElAbund}. 

%%The size of the random errors is estimated from the standard
%%deviation (1 sigma) of the abundances derived from individual lines
%%for elements that have four or more lines available for the abun-
%%dance analysis. For the abundances of elements based on less
%%than four lines, we employ the mean of the random errors esti-
%%mated from those elements with four or more lines available
%%(0.12 dex). For most of these elements, the abundances derived
%%from individual lines distribute within 0.12 dex around the mean
%%abundance that is adopted as the final result. Exceptions are Cu,
%%Dy, and Er, for which we adopt 0.20 dex as the random errors.
%%We also estimate the random errors in the abundance measure
%%ment from the uncertainties of equivalent width measurements,
%%which are estimated using the relation sigmaw ’ (kn1/2 )/(R1⁄2S/N)
%%pix(Norris et al. 2001). The errors of abundances from equivalent
%%width measurement are smaller than 0.12 dex for most lines.

\subsection{Comparison with Previous Studies}
To verify our results, we compare to the abundances measured by previous studies for the same objects. We exclude Sr from the comparison, since their abundances are less certain (\S 3.1). Differences of adopted atmospheric parameters and abundance results  (our result minus the values of previous studies) are summarized in Table \ref{tab:TEST}.
%--table7----------------------------------------------------------------------------------------
\begin{table*}[!t]
\begin{center}
\caption{ Error Estimates for BD$ +6^{\circ}648$}
\begin{tabular}{lrrp{1pt}rrp{1pt}rrp{1pt}rr}
\hline
\hline
& \multicolumn{2}{c}{$\Delta T_{\rm eff}$} & & \multicolumn{2}{c}{$\Delta$log {\it g}} & & \multicolumn{2}{c}{$\Delta$[Fe/H]} & & \multicolumn{2}{c}{$\Delta$$\xi$} \\
\cline{2-3} \cline{5-6} \cline{8-9} \cline{11-12}
   Species  & $-100$ K & +100 K & & $-0.3$ & +0.3 & & $-0.3$ & +0.3 & & $-0.3$ & +0.3 \\
\hline
\ion{Fe}{1}& $-0.19$ & 0.18 & & 0.08 & $-0.07$ & &0.02 & $-0.03$ & & 0.17 & $-0.18$ \\
\ion{Fe}{2}& $-0.03$ & 0.03 & & $-0.11$ & 0.10 & & $-0.04$ & 0.05 & & 0.08 & $-0.06$ \\
\ion{Sr}{2}& $-0.11$ & 0.13 & & $-0.01$ & 0.01 & & 0.02 & $-0.03$ & & 0.18 & $-0.17$ \\
\ion{Y}{2}& $-0.07$ & 0.06 & & $-0.08$ & 0.07 & & $-0.05$ & 0.06 & & 0.28 & $-0.24$ \\
\ion{Zr}{2}& $-0.04$ & 0.04 & & $-0.10$ & 0.10 & & $-0.06$ & 0.07 & & 0.09 & $-0.06$ \\
\ion{Mo}{1}& $-0.27$ & 0.21 & & 0.02 & $-0.02$ & & 0.08 & $-0.12$ & & 0.03 & $-0.04$ \\
\ion{Ru}{1}& $-0.29$ & 0.25 & & 0.05 & $-0.04$ & & 0.07 & $-0.09$ & & 0.02 & $-0.01$ \\
\ion{Pd}{1}& $-0.21$ & 0.21 & & 0.04 & $-0.01$ & & 0.02 & $-0.02$ & & 0.03 & $-0.04$ \\
\ion{Ba}{2}& $-0.04$ & 0.07& & $-0.09$ & 0.09 & & $-0.05$ & 0.06 & & 0.05 & $-0.07$ \\
\ion{La}{2}& $-0.05$ & 0.04 & & $-0.10$ & 0.09 & & $-0.04$ & 0.06 & & 0.09 & $-0.07$ \\
\ion{Ce}{2}& $-0.05$ & 0.05 & & $-0.10$ & 0.11 & & $-0.06$ & 0.07 & & 0.05 & $-0.03$ \\
\ion{Nd}{2}& $-0.05$ & 0.05 & & $-0.11$ & 0.10 & & $-0.06$ & 0.07 & & 0.06 & $-0.04$ \\
\ion{Eu}{2}& $-0.05$ & 0.04 & & $-0.10$ & 0.09 & & $-0.05$ & 0.06 & & 0.00 & 0.00 \\
\ion{Pb}{1}& $-0.24$ & 0.20 & & 0.07 & $-0.03$ & & 0.01 & $-0.02$ & & 0.00 & 0.00 \\
\hline 
 \end{tabular}
\label{er6648a}
\end{center}
\end{table*}
%-----------------------------------------------------------------------------------------------------------------------------

\subsubsection{BD$ +6^{\circ}648$}
Three neutron-capture elements (La, Eu, and Pb) of BD$ +6^{\circ}648$ have been studied by \citet{aoki08}. A relatively large difference of atmospheric parameters is found in the value of log $g$ (0.4 dex). According to our error estimates (Table \ref{er6648a}), such a difference in the value of log $g$ results in a change of the abundance of heavy elements around +0.09 dex. The difference in the value of $\xi$ is $-0.35$ km s$^{-1}$, which affects the abundance of La around +0.09 dex according to Table \ref{er6648a}. The estimations are comparable with the difference between our results of La and Eu, and those of \citet{aoki08}. This suggests that the differences in abundance are mostly due to the difference of the adopted atmospheric parameters. 

This object was also studied by \citet{burris00}. They have measured the abundances of Y, Zr, Ba, La, Nd, and Eu. Overall, their results are in good agreement with ours as shown in Table \ref{tab:TEST}. Largest disagreement is the measurement of Nd, for which their result is 0.12 dex higher than ours. The $T_{\rm eff}$ adopted by \citet{burris00} is lower than ours by 100 K. Such a difference in the value of $T_{\rm eff}$ results in lower abundance, while the higher log $g$ (0.20 dex) of our work partially cancels the effect.

\subsubsection{HD 23798}
Neutron-capture elements including La and Eu of HD 23798 have been studied by \citet{simm04}. We adopt $T_{\rm eff}$ from their study. The difference of $\xi$ between their study and ours is $0.32$ km s$^{-1}$, which does not significantly affect the analysis of weak lines. The abundances of the two elements are in good agreement with ours. 

\subsubsection{HD 85773}
We adopt $T_{\rm eff}$ from \citet{simm04} for this target. The atmospheric parameters are overall in good agreement. The abundances of the two neutron-capture elements La and Eu measured by them are both higher than our results by around 0.2 dex. The ratio between La and Eu (La/Eu) is comparable. The reason for the discrepancy in abundances is unclear, due to the few available details of analysis in \citet{simm04}. 

The target has also been studied by \citet{ishig13}. They measured abundances of Y, Zr, Ba, Nd, and Eu. The $T_{\rm eff}$ we adopt is $102$ K lower than their $T_{\rm eff}$. [Fe/H] derived by our analysis is also lower than that of \citet{ishig13}. Our overall abundance ratios are lower by 0.1 to 0.2 dex than their results, and the discrepancy is at least partially explained by the differences in adopted effective temperature.

\subsubsection{HD107752}
Neutron-capture elements of HD 107752 have been studied by \citet{ishig13}. They measured the abundance of Y, Zr, Ba, Nd, and Eu. The $T_{\rm eff}$ and log $g$ in our analysis are 66 K and  0.2 dex lower than those in \citet{ishig13}. According to our investigation on error estimates (Table \ref{er10a}), such differences in log $g$ and  $T_{\rm eff}$ should affect the abundance about $-0.1$ dex. The discrepancy of the abundances between the two studies is attributed to the difference in the adopted atmospheric parameters. 

\subsubsection{HD110184}
Neutron-capture elements of HD 110184 have been studied by \citet{honda04}. They measured abundances of Y, Zr, Ba, La, Ce, Nd, and Eu. The abundances of heavy neutron-capture elements from La to Eu obtained by our analysis are in good agreement with their results. The Ba abundance determined by our study is  about 0.24 dex lower than their result. This is due to the difference in the adopted lines. \citet{honda04} use the two lines at 4554 and 4934 {\AA}, which give out higher abundance compared to lines in the red range such as 5854 and 6497 {\AA} we adopted (see \S2). 
%-----table8---------------------------------------------------------
\begin{table*}[!t]
\begin{center}
\caption{Comparison with Previous Studies}
\begin{threeparttable}
\begin{tabular}{ccccccccccccc}
\hline
\hline
Star & $\Delta T_{\rm eff}$& $\Delta \mbox{log} g$ & $\Delta$ [Fe/H] & $\Delta$ $\xi$ &\multicolumn{7}{c}{$\Delta$log $\epsilon$~~(dex)} &\\
\cline{6-12}
&(K)&(dex)&(dex)&(km s$^{-1}$)&Y&Zr&Ba&La&Ce&Nd&Eu& References\\
\hline
 BD$ +6^{\circ}648$&0& $+$0.40 & $-0.01$  & $-0.35$ &\nodata  &\nodata &\nodata & $+$0.23 &\nodata &\nodata& $+$0.12 &1\\
&$-100$& $+$0.20 & $-0.01$ & $-0.15$ &$-0.09$ &$-0.07$ &$+$0.04 & $+$0.03 &\nodata &$-0.12$& $-$0.08 &2\\
HD 23798 &0 &$+$0.20 &0.00 &$-0.32$ &\nodata  &\nodata &\nodata&$+$0.02&\nodata&\nodata&$-0.14$&3\\
HD 85773& 0 & 0.00 &0.00&$-0.02$&\nodata&\nodata&\nodata&$-0.19$ &\nodata & \nodata &$-0.24$ &3 \\
& $-$102 & $+$0.14 &$-0.21$ &0.00 &$-0.22$ &$-0.20$ &$-0.02$ &\nodata &\nodata & $-0.12$ &$-0.15$ &4 \\
HD 107752&$-66$ &$-0.21$ &$-0.02$ & $+0.05$ &$-0.02$&$-0.19$&$-0.19$ &\nodata &\nodata &$-0.07$&$-0.06$&4\\
HD 110184 & 0&$+$0.20&0.00 & $+$0.30 &$-0.42$ & $-0.32$ &$-0.24$ &$+$0.03 & $+$0.19&$+$0.03&$-0.12$&5\\
\hline
\end{tabular}
\tablerefs{(1)\citet{aoki08}, (2) Burris et al. (2000), (3) Simmerer et al. (2004), (4) Ishigaki et al. (2013), (5) Honda et al. (2004)}
\tablecomments{The difference is taken in a sense of our results minus other works 
}
\end{threeparttable}
\label{tab:TEST}
\end{center}
\end{table*}
%--------------------------------------------------------------------------------------------------------------

Our results of Y and Zr have differences of around 0.4 dex from theirs. The discrepancy is also due to the difference in the adopted lines. \citet{honda04} use five lines to determine the abundance of Y. Of these five lines, only 3950 {\AA} is included in the lines we adopt. The equivalent width of 3950 {\AA} measured by our study is 89.7 m{\AA} and abundance (log$\epsilon$(Y)) derived is $-0.53$ dex. \citet{honda04} measured 88.7 m{\AA} for this line, which is consistent with our measurement. Other lines used by \citet{honda04} are not included in our study, because some are within the gap of our spectrum and some are suspected to be affected by contamination. The abundances obtained from other lines we adopted are lower than that derived from the 3950 {\AA} line, which renders our result, which is lower than that in \citet{honda04}. 

The abundance of Zr derived by \citet{honda04} is 0.32 dex higher than our result. We adopt six lines to determine the abundance of Zr, whereas \citet{honda04} adopted three lines (4161, 4209, and 4317 {\AA}). Among these lines, we do not adopt 4161 {\AA} due to contamination suspected for the line. The average of the abundance obtained from the other two lines (4209 and 4317 {\AA}) is 0.20 dex, which is in good agreement with \citet{honda04}, whereas the abundance derived from the rest of the lines adopted in our study is higher than theirs. 

The abundances of Y, Zr, Pd, Ba, and Eu are also studied by \citet{hans12}. In addition, \citet{hans14} measured Mo and Ru for this object. The abundances ($\log \epsilon $ values) of these elements derived by Hansen et al. (2012, 2014) are 0.5 -- 0.8 dex higher than our results. This is partially explained by the difference of adopted atmospheric parameters: the $T_{\rm eff}$ we adopt is 210~K lower than theirs. Differences of abundances between the two studies, however, remain to be of the 0.2~dex level, even after the differences of parameters are taken into consideration. The reason for the discrepancy is unclear, due to the few available details of abundance analysis for this object in their studies. It should be noted, however, that these differences are systematic, and the overall abundance patterns obtained by the two studies are similar to each other.

\section{Results and Discussion}
\subsection{Comparison with the Solar r-process Pattern}\label{Comp-chap} 
\begin{figure*}[!t]
\centering
\subfigure{
        \resizebox{80mm}{!}{\includegraphics{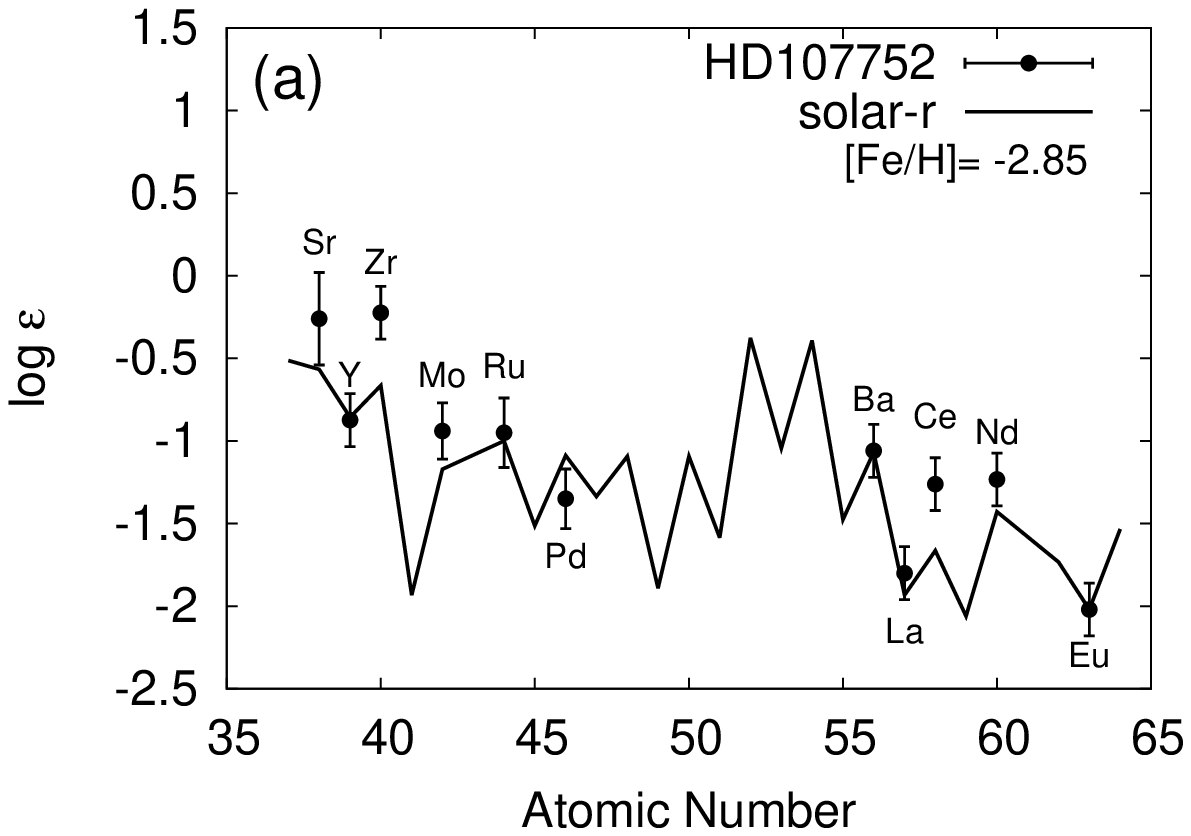}}
\hspace{1mm}
        \resizebox{80mm}{!}{\includegraphics{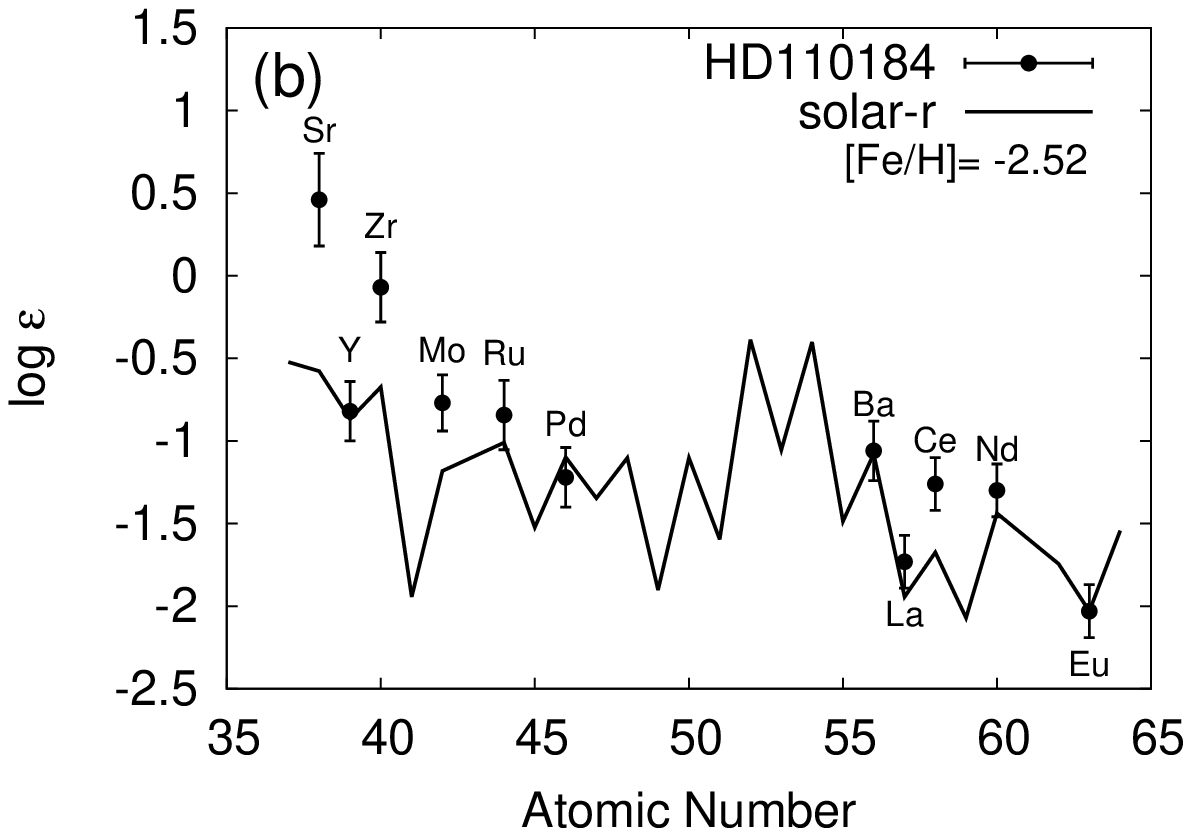}}
}
\subfigure{
        \resizebox{80mm}{!}{\includegraphics{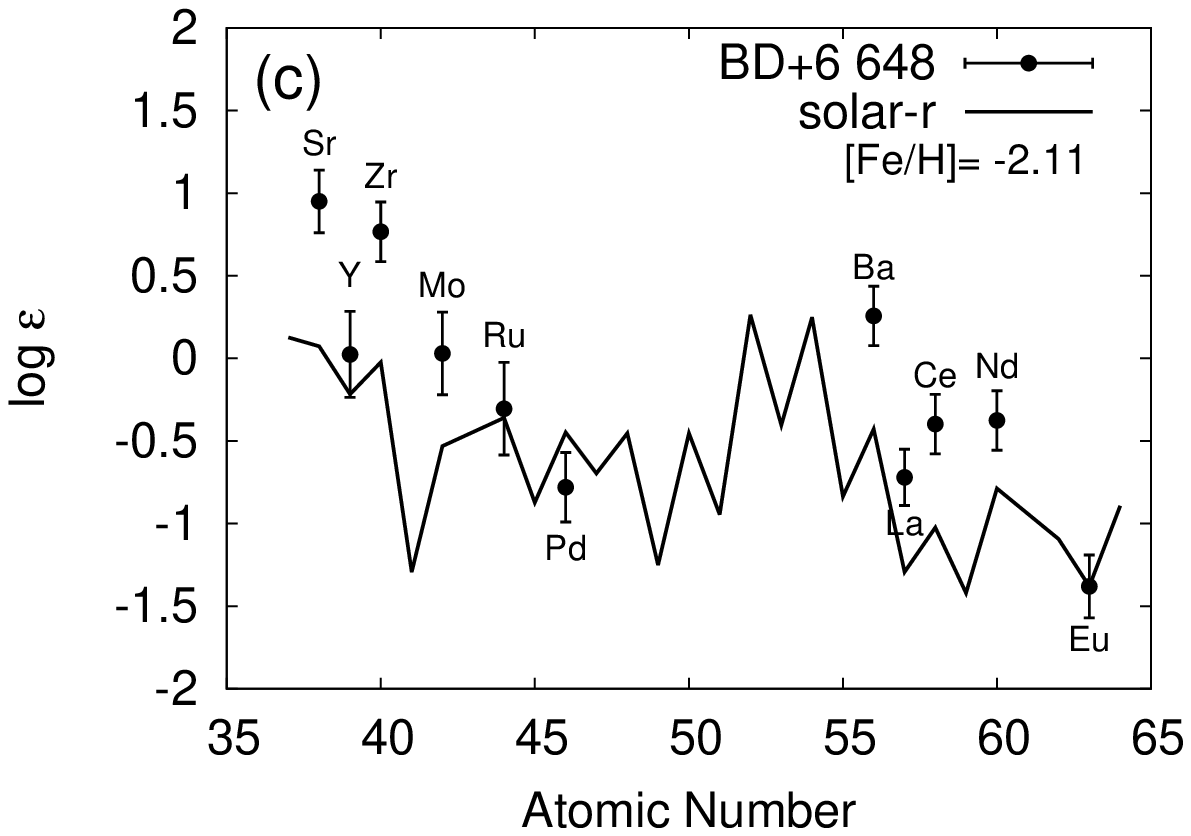}}
\hspace{1mm}
        \resizebox{80mm}{!}{\includegraphics{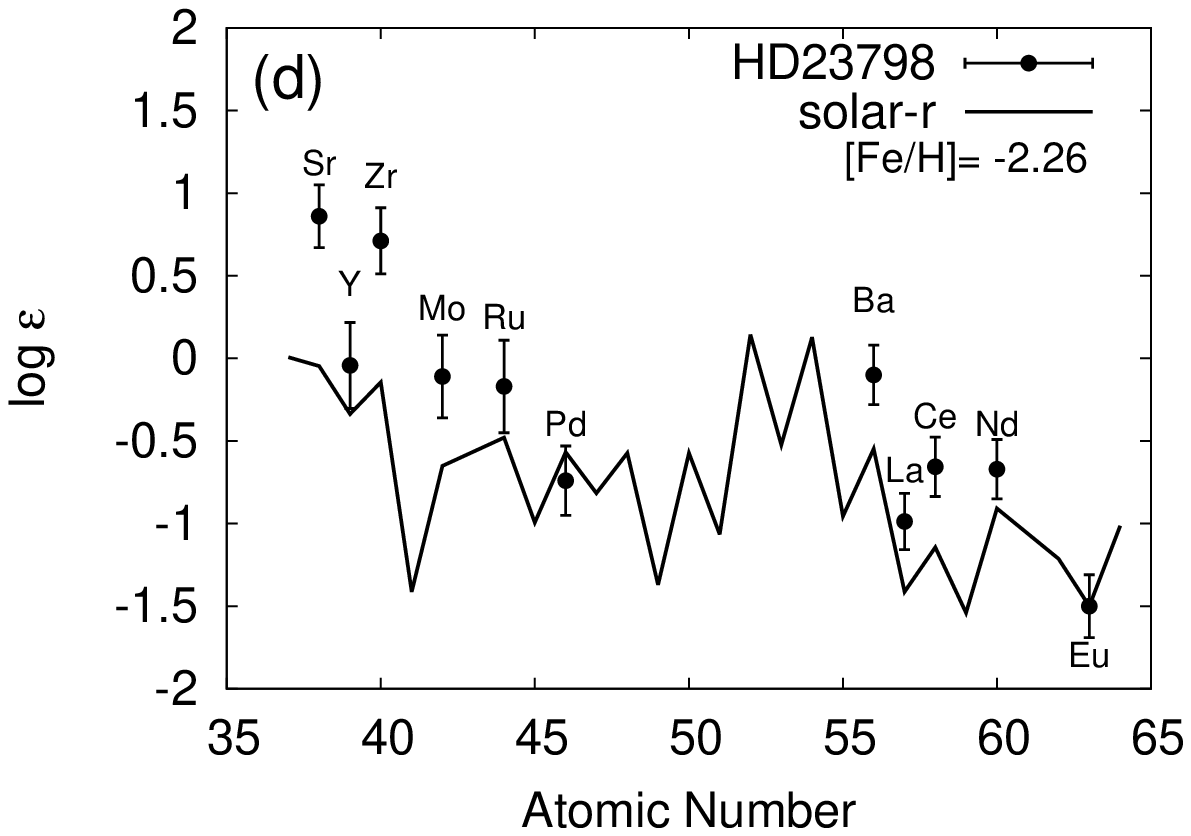}}
}
\subfigure{
        \resizebox{80mm}{!}{\includegraphics{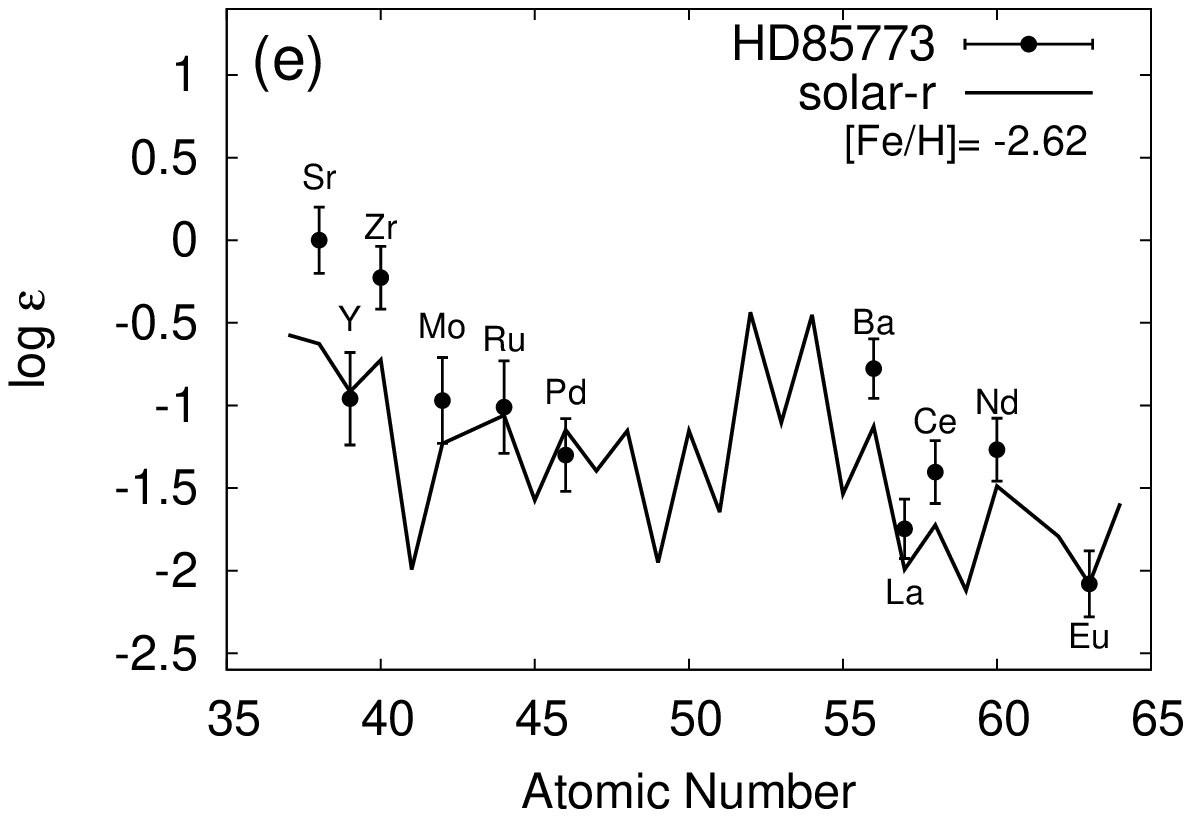}}
\hspace{1mm}
        \resizebox{80mm}{!}{\includegraphics{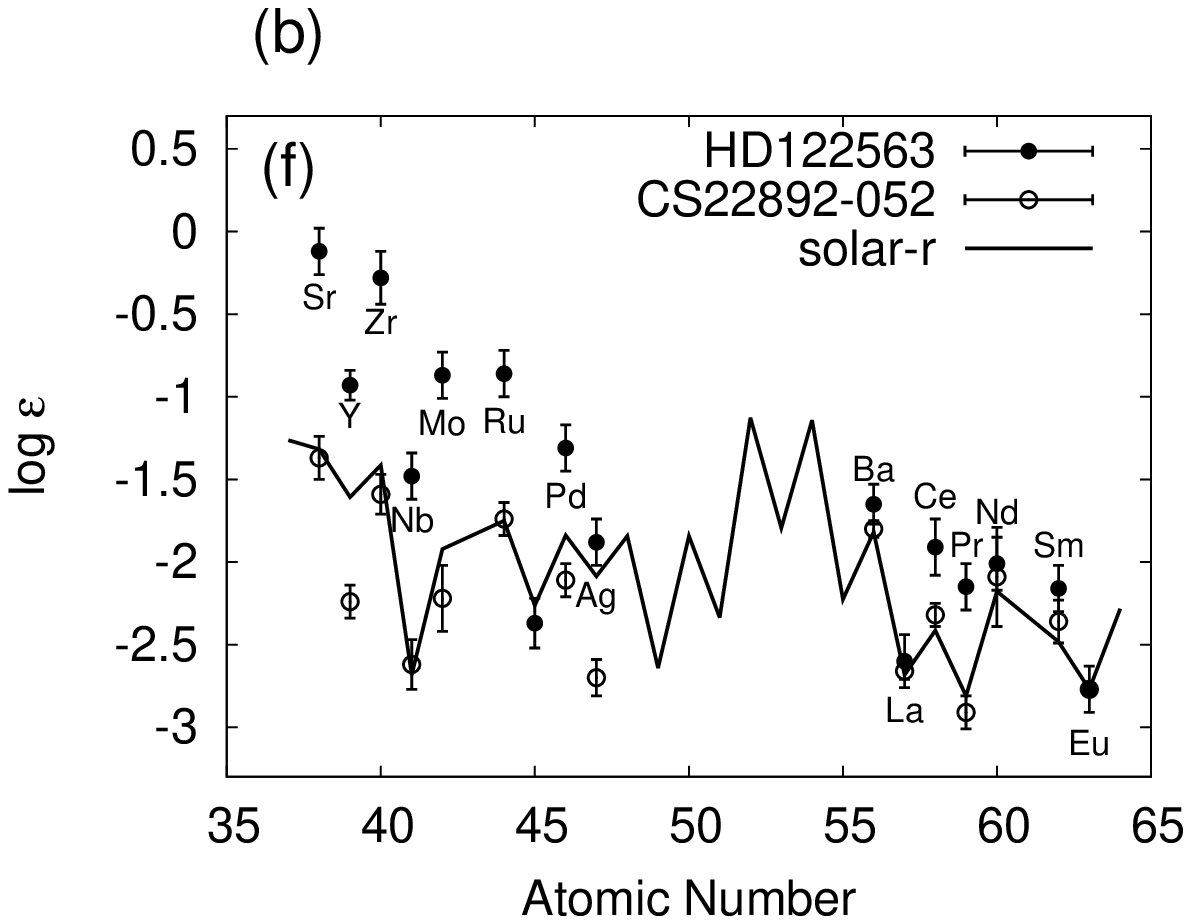}}
}
\caption{Abundances of neutron-capture elements in five target stars. The abundance patterns are compared to the solar-system r-process pattern (normalized at Eu). Panel (f) shows the abundance patterns of HD~122563 in black circles and CS 22892-052 in open circles.}
\label{abund}
\end{figure*}

Figures \ref{abund} (a) - (e) show abundances of the five target stars compared to the Eu scaled solar r-process pattern \citep{simm04}. The abundance pattern of HD~122563 \citep{honda06} and CS 22892-052 \citep{sneden03} are also compared to the solar r-process pattern in Figure \ref{abund} (f). The abundance patterns of HD 122563 and the solar r-process are scaled to the Eu abundance of CS 22892-052.

%------figure4------------------------------------------------------------------------------------------
%-----------------------------------------------------------------------------------------------------------------------------
As shown in Figure \ref{abund} (a) and (b), the abundance patterns of heavy neutron-capture elements of the two stars, HD 107752 and HD 110184, agree with that of the scaled solar r-process. An exception is Ce, which is overabundant at about 0.4 dex compared to the scaled solar r-process pattern in both stars. Similar disagreement is also found in the abundance of HD~122563; its measurement of Ce results in the overabundance of about 0.5 dex (Figure \ref{abund} (f)). 

On the contrary, the abundances of light neutron-capture elements of all the target stars tend to be overabundant compared to the solar r-process pattern normalized at Eu. The levels of overabundance are not as evident as found in HD~122563 (Fig. \ref{abund} (f)). This is expected from our target selection (Fig. \ref{zrba}), since our targets are known to have higher [Zr/Ba] ratios than the solar values, but not as extreme as HD~122563. Such an overabundance of light neutron-capture elements suggests contribution of the weak r-process. In particular, the abundances of Sr and Zr are significantly overabundant. An exception is Y, which shows no excess compared to the scaled solar r-process pattern. Similar trend of Y is also found in CS 22892-052 and HD~122563 (Fig. \ref{abund}(f)). Mo is overabundant in all the objects. However, not all of the stars show an overabundance in Ru, while Pd is under-abundant in all of the objects. 

The three target stars BD$ +6^{\circ}648$, HD 23798, and HD 85773 (Fig.~\ref{abund} (c)--(d)) also show overabundances of heavy neutron-capture elements, especially of Ba. We discuss possible contributions of the s-process to these elements in section \S4.2.

\subsection{Contributions of Main r-process, Weak r-process and s-process}
\begin{figure*}[!t]
\centering
\subfigure{
        \resizebox{80mm}{!}{\includegraphics{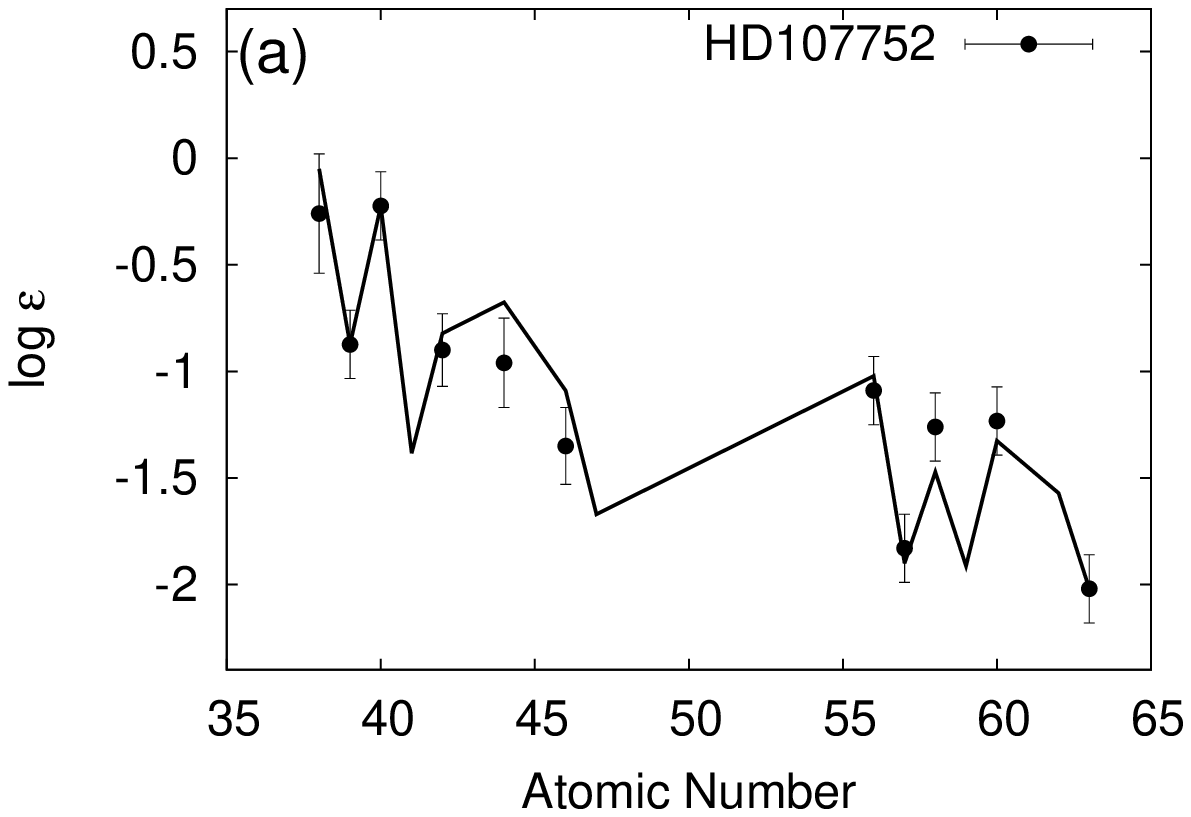}}
\hspace{1mm}
        \resizebox{80mm}{!}{\includegraphics{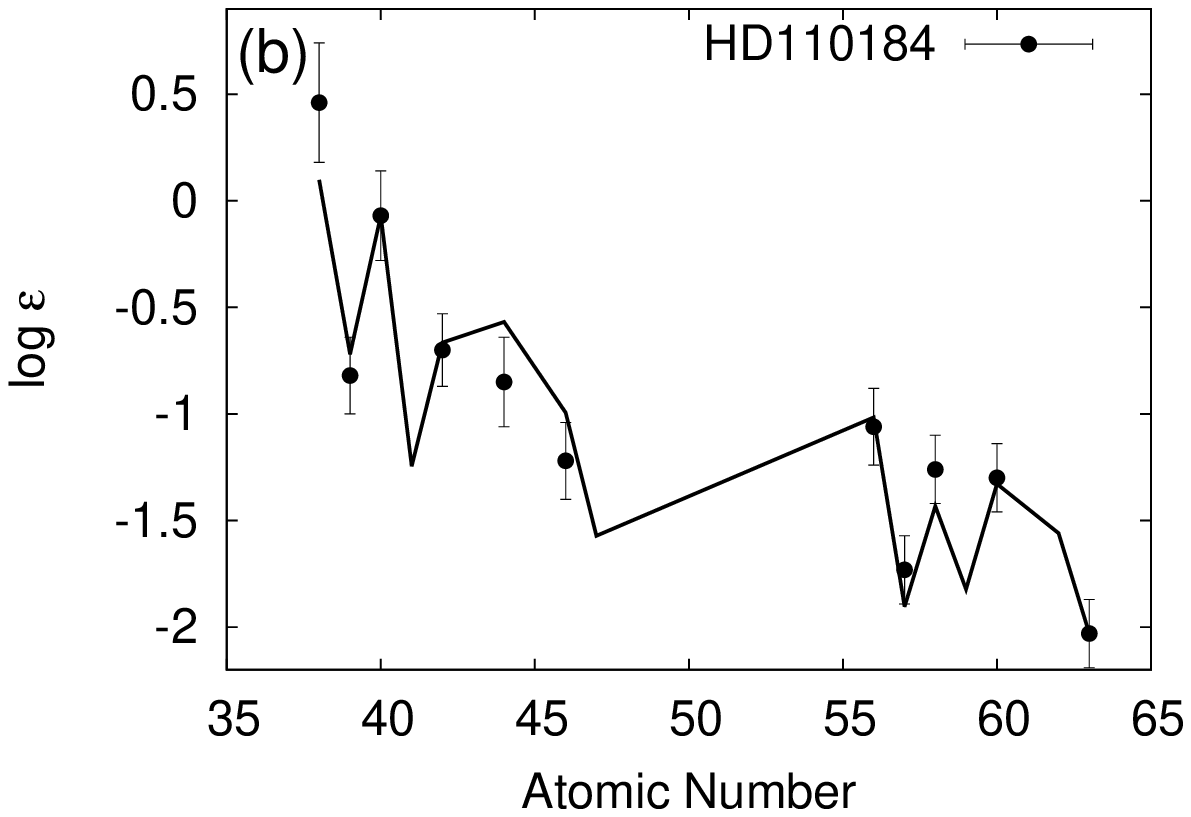}}
}
\subfigure{
        \resizebox{80mm}{!}{\includegraphics{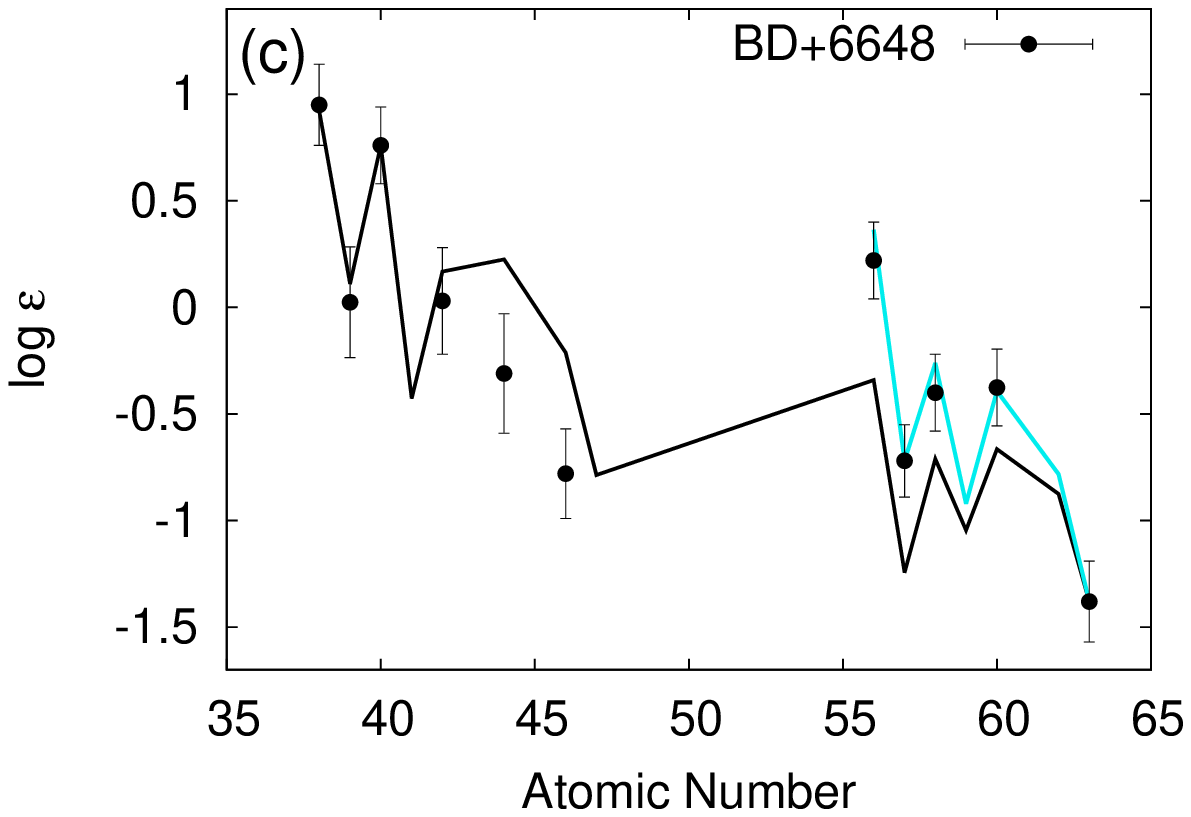}}
\hspace{1mm}
        \resizebox{80mm}{!}{\includegraphics{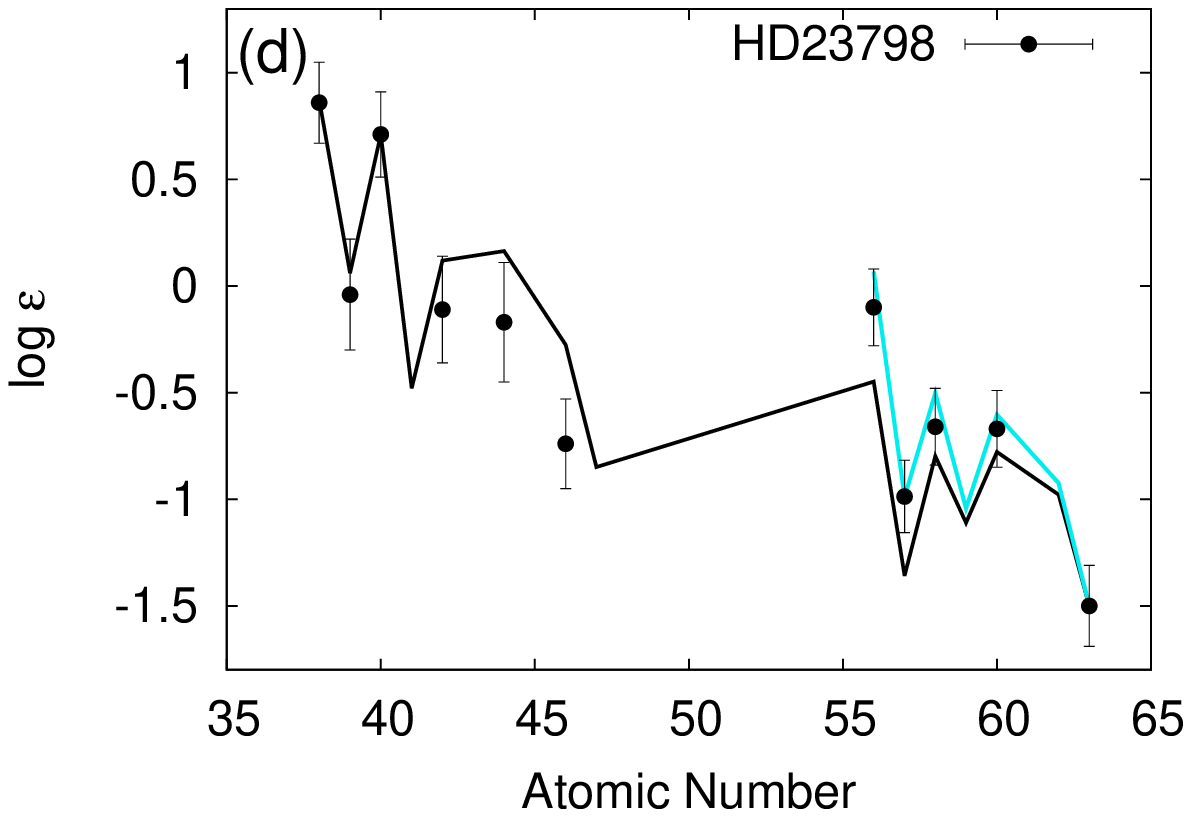}}
}
\subfigure{
        \resizebox{80mm}{!}{\includegraphics{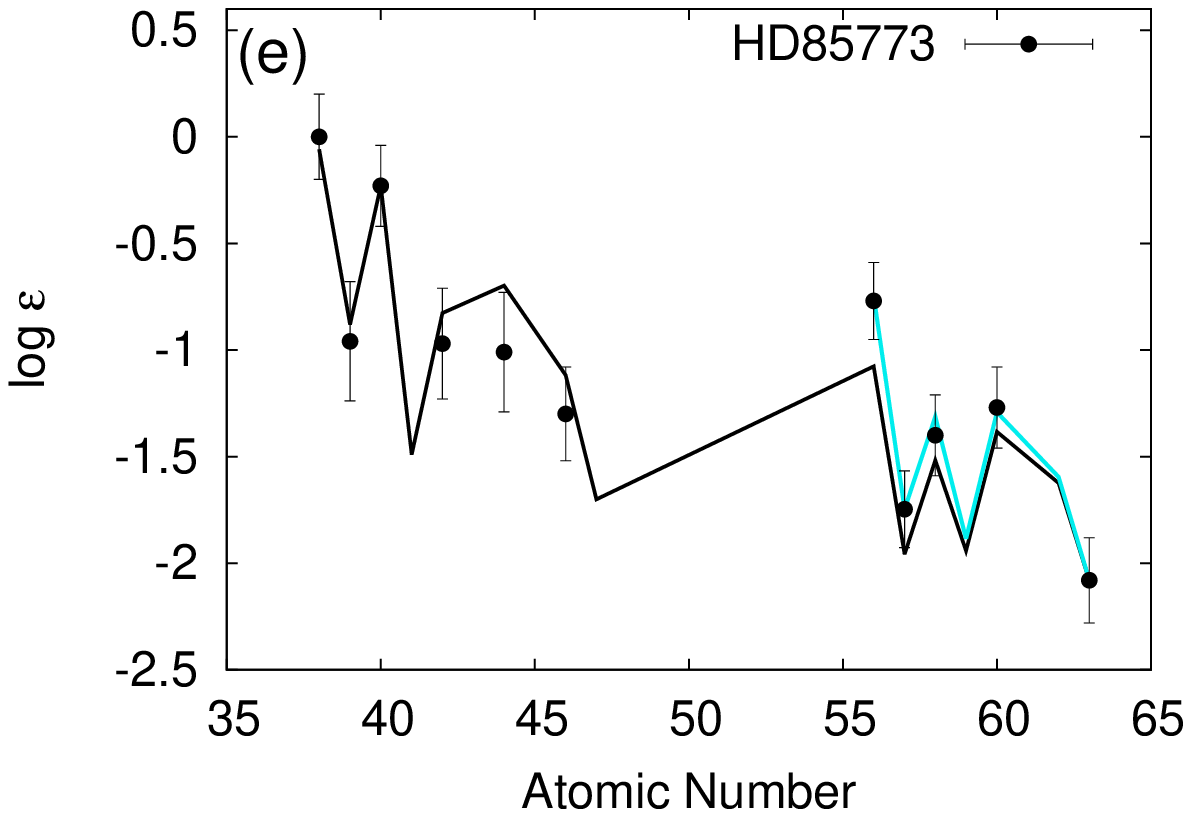}}
\hspace{1mm}
        \resizebox{80mm}{!}{\includegraphics{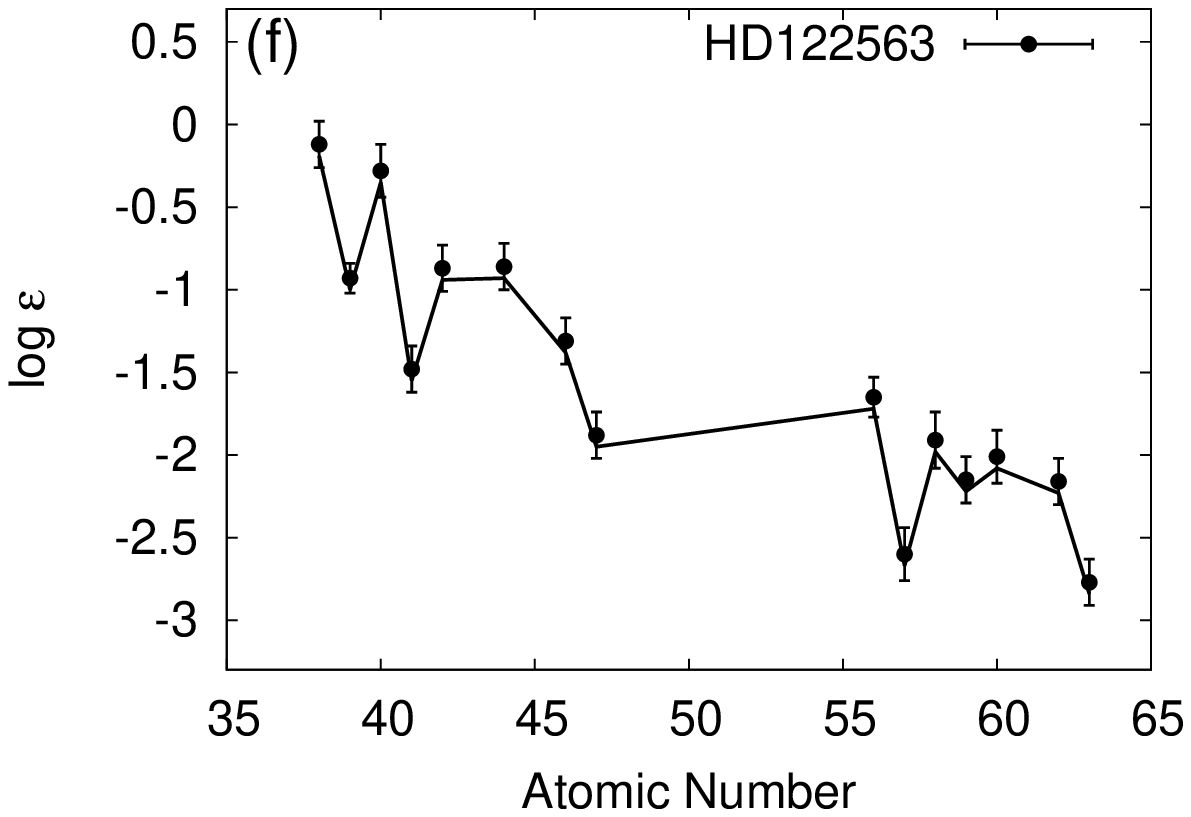}}
}
\caption{Abundances of neutron-capture elements in HD 122563 and five target stars compared to the mixed abundances (solid black line) of the main r-process (CS 22892-052) and the weak r-process (HD~122563). The abundance of  heavy neutron-capture elements is compared to the solar s-process abundance, which is added to the mixed abundance pattern to fit the abundance of La and Eu.}
\label{percent2}
\end{figure*}

We interpret the excess of light neutron-capture elements in our objects as contamination of the weak r-process. If this is the case, the abundances of our target stars will be explained by the average of the abundances of the main and weak r-processes with certain weights, assuming that both processes have universal patterns. We attempt to explain the abundance distributions of our target stars by adopting the abundances of CS 22892-052 and HD~122563 as representative of the main r-process and the weak r-process, respectively. The contributions of the two components are determined to reproduce the abundance ratios between Zr and Eu of each target star. We calculate the weights of the main r-process ({\it f}) and the weak r-process ($1-f$) as 
\begin{equation}
\bigl(f\varepsilon_{m,\mbox{\scriptsize Zr}}+(1-f)\varepsilon_{w,\mbox{\scriptsize Zr}}\bigr):\bigl(f\varepsilon_{m,\mbox{\scriptsize Eu}}+(1-f)\varepsilon_{w,\mbox{\scriptsize Eu}}\bigr)= \varepsilon_{\star,\mbox{\scriptsize Zr}}:\varepsilon_{\star,\mbox{\scriptsize Eu}}, 
\end{equation} 
where $\varepsilon_{m,i}$ and $\varepsilon_{w,i}$ are the abundances of element $\it{i}$ for CS 22892-052 and HD~122563, respectively. $\varepsilon_{\star,i}$ is the abundance of element $i$ for each of target star. The parameter $f$ is the contribution weight of the main r-process to Zr for each star, as presented in Table \ref{tab:ratio}. We apply the weight of main r-process contribution to other elements by calculating
\begin{equation}
\varepsilon'_{i}=f\varepsilon_{m,i}+(1-f)\varepsilon_{w,i}
\end{equation} 
where $\varepsilon'_{i}$ is the linear abundance of each element. Table \ref{tab:ratio} also gives the weak r-process contribution ratio calculated for Zr
\begin{equation}
R_{w}\mbox{(Zr)}=\frac{(1-f)\varepsilon_{w,\mbox{\scriptsize Zr}}}{f\varepsilon_{m,\mbox{\scriptsize Zr}}+(1-f)\varepsilon_{w,\mbox{\scriptsize Zr}}}
\end{equation} 
Figures \ref{percent2}(a)-(e) show our target stars compared with the mixtures of the main and weak r-process patterns with contribution weights of the main r-process presented in Table \ref{tab:ratio}. The lines fit well to the abundance distributions of the stars HD 107752 and HD 110184 (Fig. \ref{percent2} (a) and (b)).
%--figure5-----------------------------------------------------------------------
%-----------------------------------------------------------------------------------------------------------------------------

Figures \ref{percent2} (c)--(e) show results of BD$ +6^{\circ}648$, HD 23798, and 
%--table9---------------------------------------------------------------------------------------
\begin{table}[!b]
\begin{center}
\caption{Main r-process, weak r-process, and s-process contribution ratio}
\begin{threeparttable}
\begin{tabular}{ccccc}
\hline
\hline
Object Name & $f$ &$R_{w}$(Zr)  &$f_{s}$ &$R_{s}$(La) \\
\hline
    BD$ +6^{\circ}648$ &0.02 &  0.94 & 0.99 &  0.71\\
    HD 23798 & 0.01 & 0.95  &0.99  & 0.58 \\
    HD 85773  & 0.06 & 0.83 & 1.00 & 0.39 \\
    HD 107752 & 0.07 & 0.80  & \nodata & \nodata  \\
    HD 110184 & 0.04 & 0.87 & \nodata & \nodata \\
\hline
 \end{tabular}
\end{threeparttable}
\label{tab:ratio}
\end{center}
\end{table}
%-----------------------------------------------------------------------------------------------------------------------------
HD 85773, respectively. Contrary to the cases for HD 107722 and HD 110184, some light elements such as Ru cannot be reproduced by the mixed abundance patterns. Furthermore, already pointed out in \S 4.1, some heavy neutron-capture elements such as Ba and La are overabundant compared to Eu in these three stars. In Figure \ref{percent2} (c) -- (e), we add cyan lines, which represent contribution of the s-process to heavy neutron-capture elements. For the s-process, the contribution weights are calculated in the same manner applied to the contribution of the weak r-process, so that the abundance ratios of La to Eu match the observational results. We adopted the solar s-process pattern by \citet{simm04}. The s-process weight ($f_{s}$) and the s-process contribution to La, $R_{s}$(La), are presented in Table \ref{tab:ratio}. The abundance patterns of heavy neutron-capture elements for these three stars are well reproduced by adding s-process (Fig. \ref{percent2}), suggesting their overabundance is due to the s-process contamination. \citet{honda06} also found such overabundance in heavy neutron-capture elements of HD 122563 and investigated for the contribution of the main s-process. They searched for a combination of the solar r-process and s-process pattern that gives the best fit to the abundance patten of the star. They concluded that some heavy neutron-capture elements such as Ba and La have little contribution of s-process, and thus have no influence on the light neutron-capture elements.

\begin{figure*}[!t]
\epsscale{1.0}
\plottwo{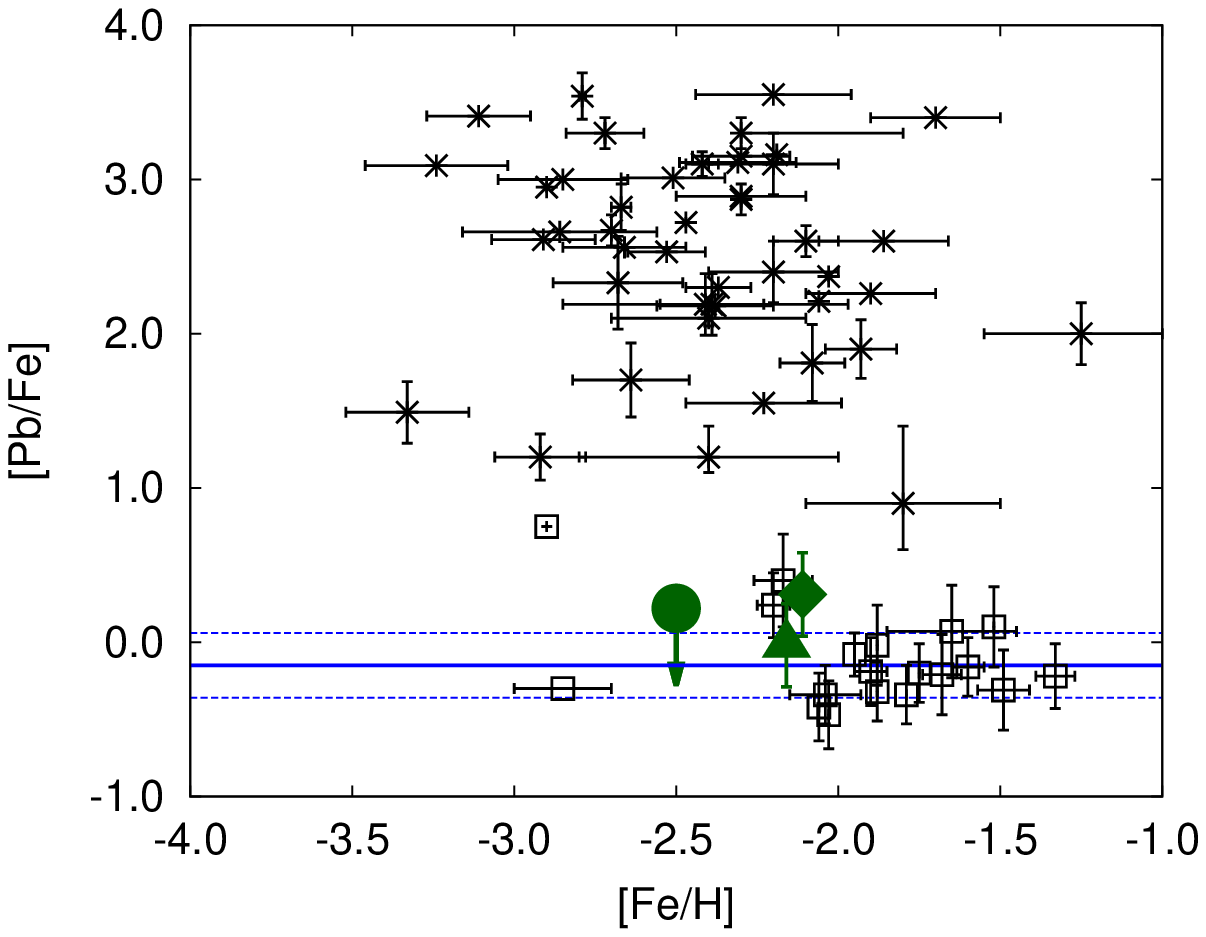}{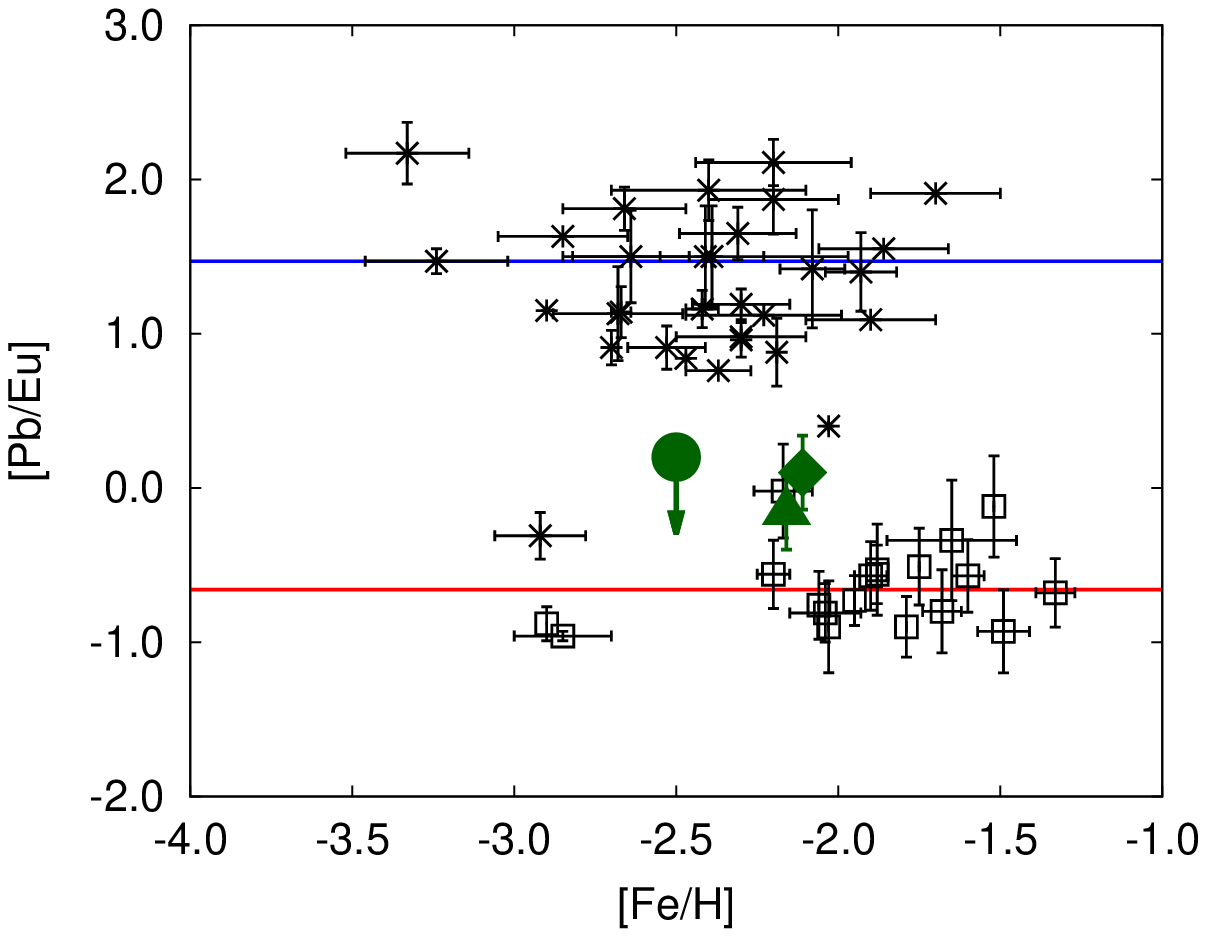}
\caption{Abundance ratios of [Pb/Fe] (left) and [Pb/Eu] (right) as functions of [Fe/H]. Black asterisks show data of CEMP stars. Black open squares show metal-poor stars taken from the SAGA database. Our results are shown by the large diamond for BD$ +6^{\circ}648$; large triangle for HD 23798; large filled circle with upper limit for HD 85773. The blue solid line at [Pb/Fe] (left) shows the average abundance ratio of metal-poor stars from previous studies. The blue dotted lines show the average standard deviation. The upper blue and lower red solid lines at [Pb/Eu] (right) show the abundance ratios of the solar-system s-process and r-process components, respectively}
%%The blue dotted lines show the abundance ratios of the solar-system r-process components (Simmerer et al. 2004).
\label{pbfe}
\end{figure*}

We also inspect the s-process contributions from abundances of Pb. At low metallicity, Pb is sensitive to s-process nucleosynthesis \citep[e.g.,][]{bus99}. We measure the Pb abundance of BD$ +6^{\circ}648$ and HD 23798, and estimate the upper limit for HD 85773.  Figure \ref{pbfe}(left) shows the [Pb/Fe] abundance
%-figure6------------------------------------------------------------------------
%----------------------------------------------------------------------------------------------------------------------------- 
ratios as a function of [Fe/H] with plots of previous studies \citep{sneden98, hill02, ivans06, aoki08, roed09, roed10, roed12b}, with our three objects. Most Carbon-Enhanced Metal-Poor stars (CEMP) with s-process abundances have higher [Pb/Fe] than normal metal-poor stars possibly due to mass transfer from companion AGB stars. The solid line of Figure \ref{pbfe} shows the average abundance ratio ([Pb/Fe] $\sim  -0.2$ dex) of normal metal-poor stars taken from previous studies, and the dotted lines show the standard deviations around the average values. Our objects have relatively high [Pb/Fe] compared to the average abundance ratio of normal metal-poor stars. Figure \ref{pbfe}(right) shows the abundance ratios of [Pb/Eu]. Assuming that Eu represents r-process elements, the abundance ratio of Pb to Eu is an indicator of s-process contamination. The upper and lower solid lines show the abundance ratios of the solar-system s-process and r-process components, respectively \citep[]{sneden96}. The ratio of the solar r-process component is comparable to the average abundance ratio of normal metal-poor stars ($\sim -0.63$ dex). Apparently, the [Pb/Eu] ratios of the two objects in our sample are also slightly higher than the average abundance ratio of normal metal-poor stars. 

These results suggest small contributions of the s-process to BD$ +6^{\circ}648$ and HD 23798. However, the light neutron-capture elements are considered to be less produced by the s-process. In low metallicity, the number of neutrons per seed nuclei is larger, resulting in efficiently producing the heavier neutron-capture elements \citep[e.g.,][]{bus99}. We interpret that the excess of light neutron-capture elements in these stars, as well as in HD 107752 and HD 110184, is attributed to contributions of the weak r-process.

\subsection{Detailed Abundance Pattern of Light Neutron-capture Elements}
As discussed in Section 4.2, the light neutron-capture elements of our target stars can be interpreted at least partially as the products of the weak r-process. In this subsection, 
%--figure7---------------------------------------------------------------------------
\begin{figure}[!h]
\epsscale{1.0}
\plotone{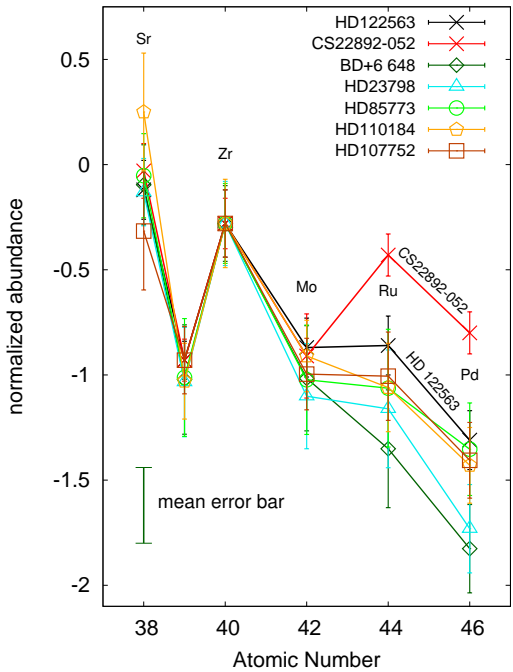}
\caption{Abundance patterns of light neutron-capture elements from Sr to Pd for the five observed stars, HD 122563 and CS 22892-052. The patterns are scaled at the Zr abundance of HD 122563 (black line).}
\label{lighte}
\end{figure}
%-----------------------------------------------------------------------------------------------------------------------------
%----figure8----------------------------------------------------------------------------------
\begin{figure}[!h]
\epsscale{1.0}
\plotone{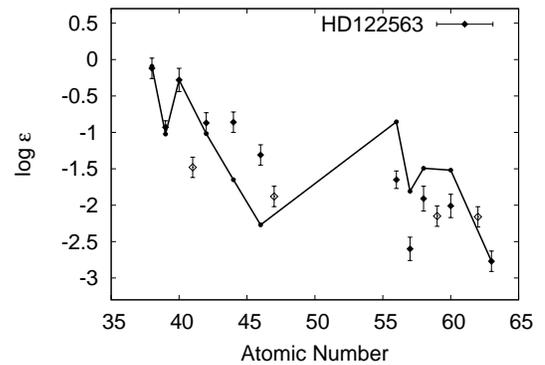}
\caption{Abundances of neutron-capture elements in HD 122563 compared to the mixed abundances of the main r-process (CS 22892-052) and BD$ +6^{\circ}648$. }
\label{BD-CSperc}
\end{figure}
%-----------------------------------------------------------------------------------------------------------------------------
we discuss the nucleosynthetic nature of the weak r-process from our observational results. We compare the abundances of light neutron-capture elements to inspect whether the weak r-process yields a universal pattern as found in the case of the main r-process.

Figure \ref{lighte} shows the abundances of light neutron-capture elements from Sr to Pd for the five observed stars. The abundances of HD 122563 and CS22892-052 are also shown for comparison purposes. The abundance patterns for the five stars are scaled to match the Zr abundance of HD 122563, since Zr can be a representative element of the weak r-process and its abundance is better determined than the Sr abundance.

As found in Figure \ref{lighte}, the abundance patterns of HD 107752, HD 110184 and HD 85773 agree within the measurement errors. Compared to these stars, the abundances patterns of HD 23798 and BD$ +6^{\circ}648$ drastically drop toward Pd. Clearly, these drops cannot be reproduced by a combination of the abundance patterns of CS 22892-052 and HD 122563. 
%%If we assume that the light neutron-capture elements of HD 122563 represent the pure abundance pattern produced by the weak r-process, the abundances of our target stars cannot be reproduced by the combination of the patterns of CS 22892-052 and HD~122563.

The steeply descending abundance patterns in BD$ +6^{\circ}648$ and HD 23798 might indicate that these stars better represent the pure weak r-process pattern rather than HD 122563. Therefore, we examine whether the combination of the abundance patterns of CS 22892-052 and BD$ +6^{\circ}648$ reproduces that of HD 122563. Figure \ref{BD-CSperc} shows the abundances of Figure \ref{BD-CSperc} shows the abundances of HD 122563 compared to the 
%---figure9-----------------------------------------------------------------------------------
\begin{figure*}[!t]
\centering
\subfigure{
        \resizebox{93mm}{!}{\includegraphics{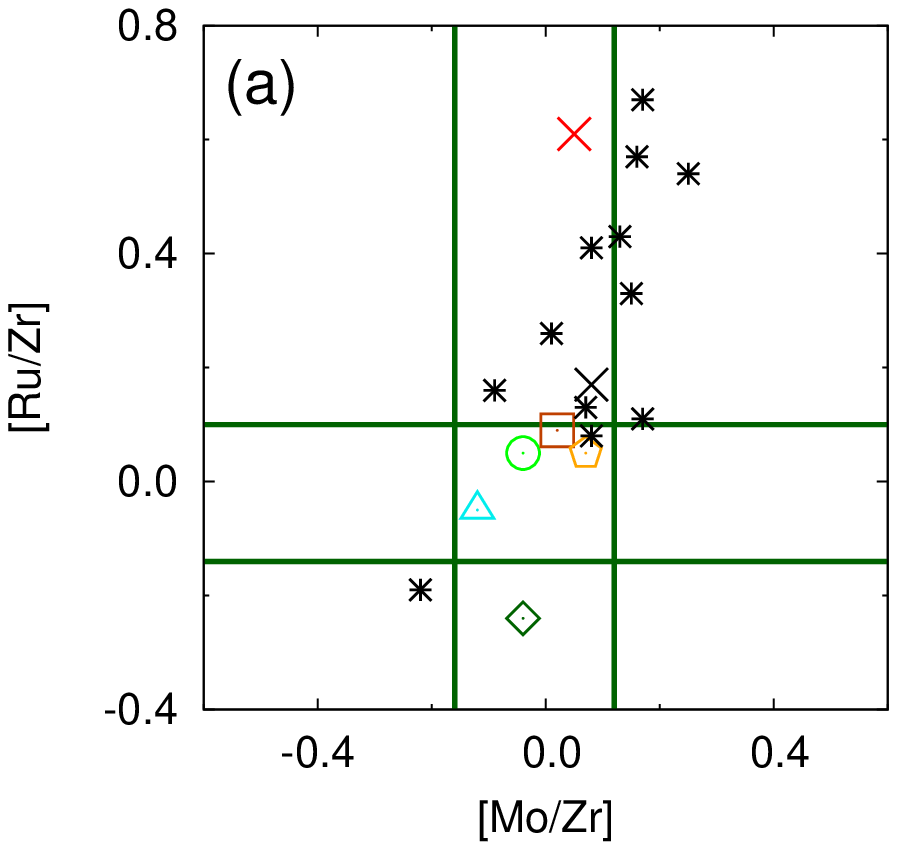}}
\hspace{-15mm}
        \resizebox{93mm}{!}{\includegraphics{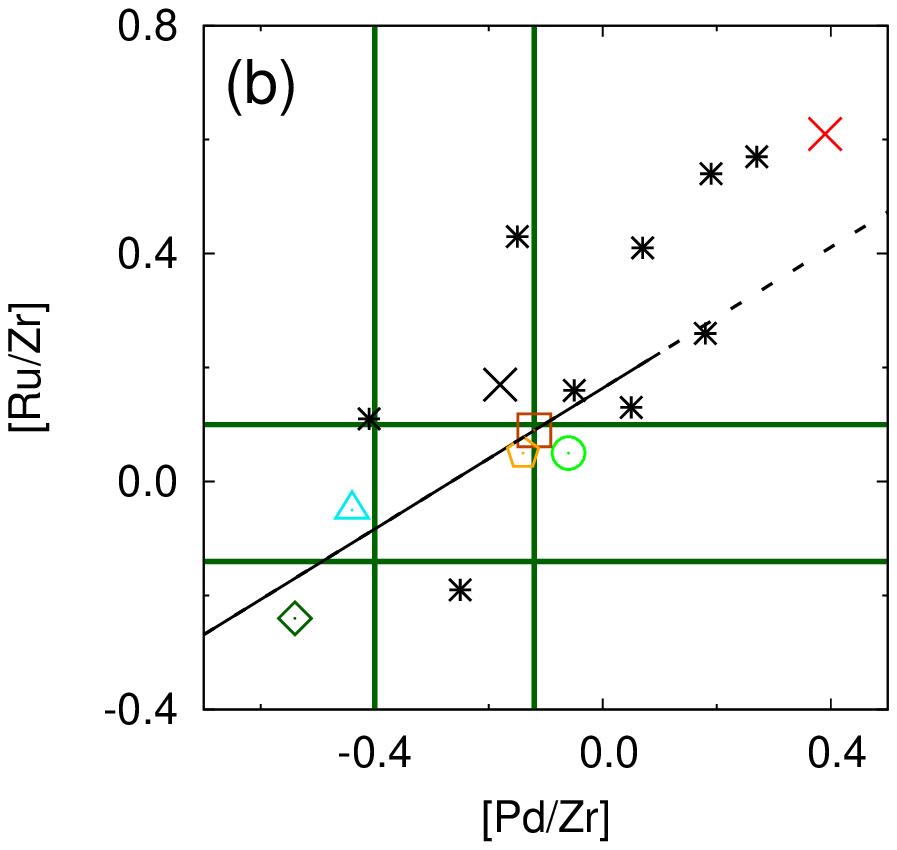}}
}

\caption{ [Ru/Zr] abundance ratios as functions of (a) [Mo/Zr] and (b) [Pd/Zr]. The symbols are the same as in Figure \ref{lighte}. The black asterisks show the selected metal-poor giants ([Fe/H]$<-2.0$) studied by \citet{hans14}. The green lines show the range of random error for each abundance ratio of our five target stars. The solid line on the panel (b) shows the linear regression lines of five analyzed stars and HD~122563. The extended dotted line shows that CS 22892-052 (red cross) is also along the regression lines.}
\label{ZrMoRuPd}
\end{figure*}
%---------------------------------------------------------------------------------------------------------------------------
solid line indicating a combination of the abundance patterns of CS22892-052 and BD$ +6^{\circ}648$ at certain weights, in the same manner as we did for Figure \ref{percent2}. As found in Figure \ref{BD-CSperc}, the heavy neutron-capture elements expected from the mixture are much more abundant than those of HD 122563, because a significant contribution of the main r-process is required to reproduce the abundance pattern of light neutron-capture elements for this object. In addition, the mixed pattern fails to reproduce the Ru and Pd abundances of HD 122563. Hence, we conclude that the light neutron-capture elements are produced by the weak r-process, and their abundances have diversity.
%-----figure10---------------------------------------------------------------------------------------
\begin{figure*}[!t]
\centering
\epsscale{.90}
\subfigure{
        \resizebox{62mm}{!}{\includegraphics{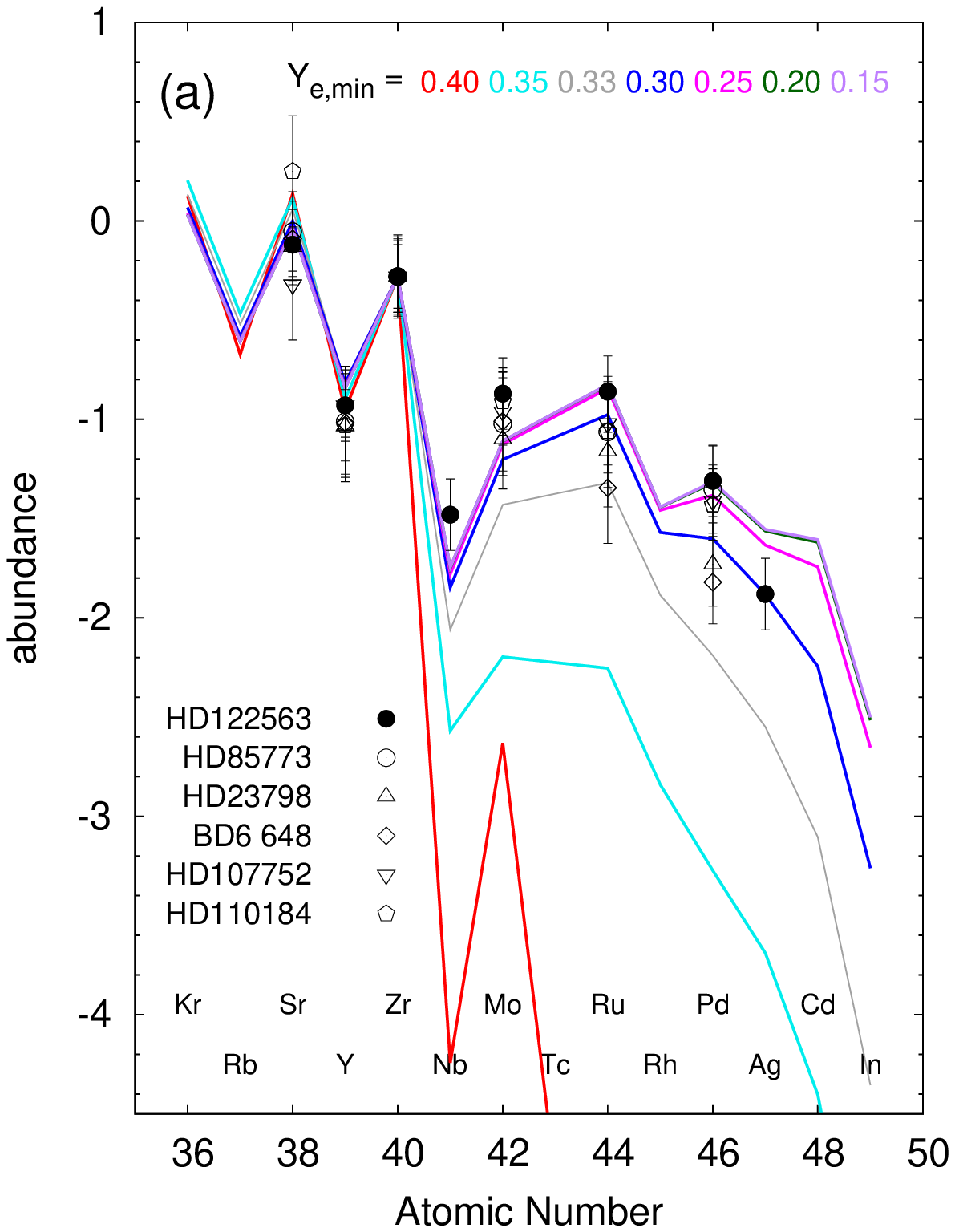}}
\hspace{-5mm}
        \resizebox{62mm}{!}{\includegraphics{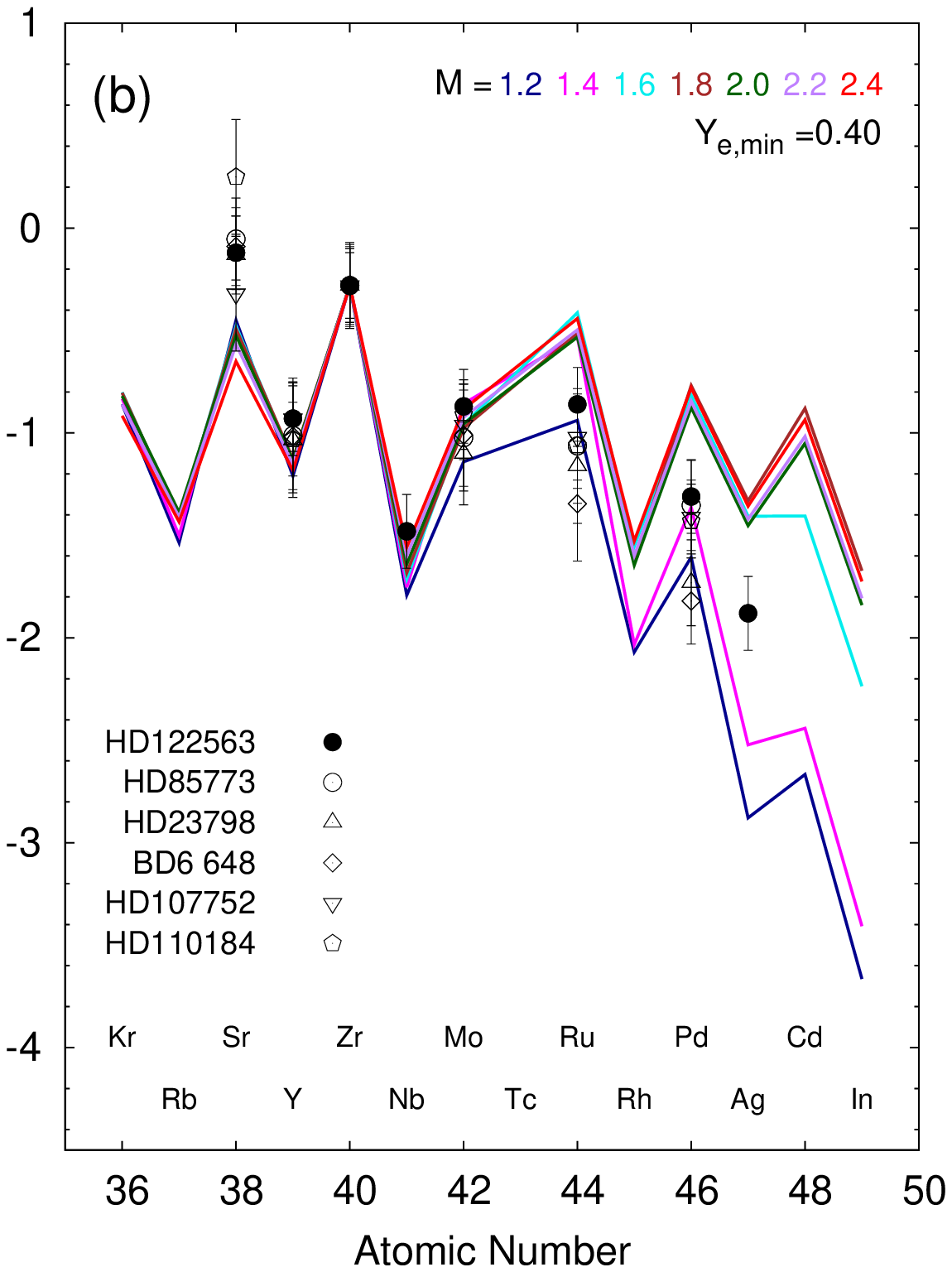}}
\hspace{-5mm}
        \resizebox{62mm}{!}{\includegraphics{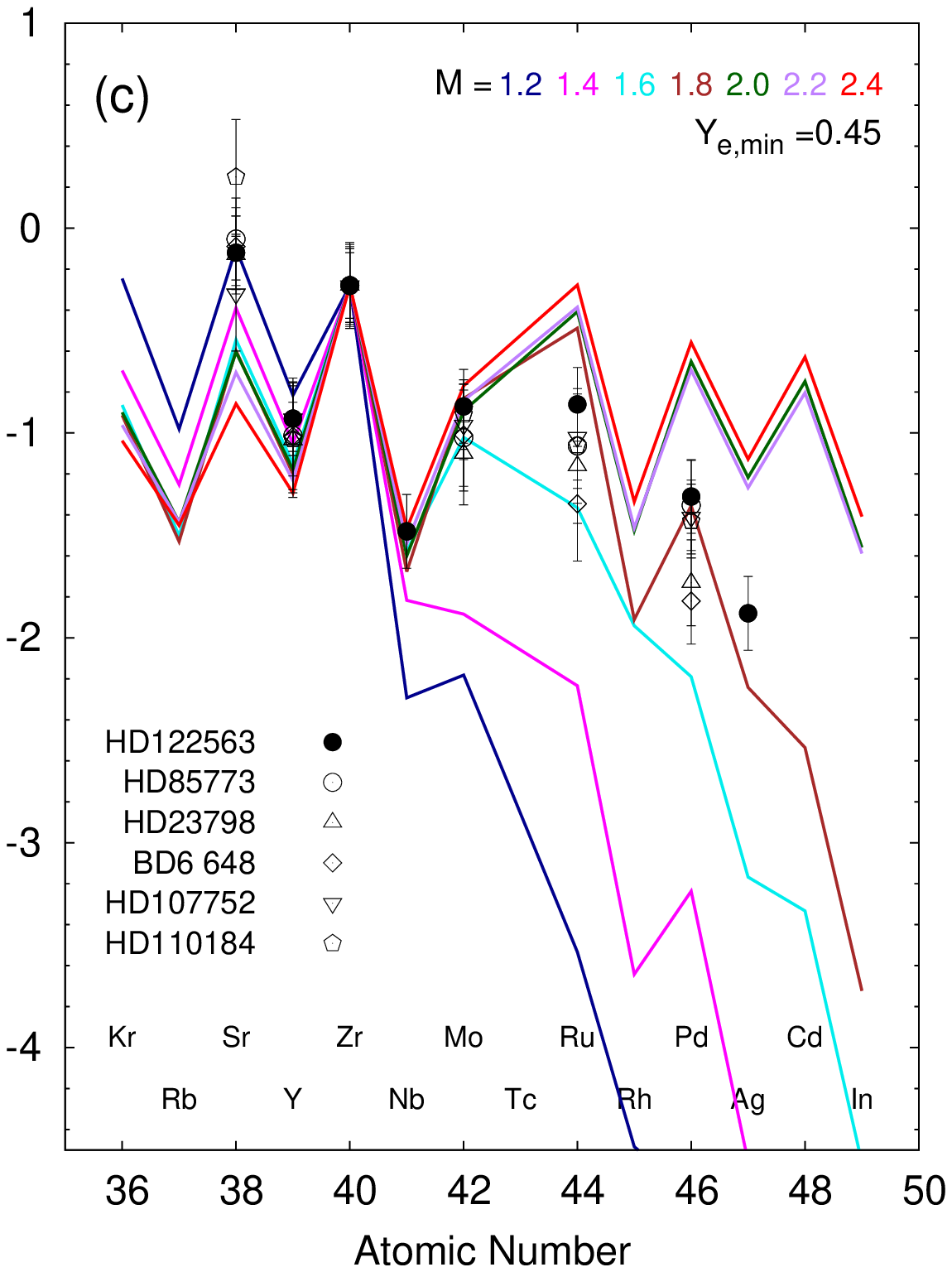}}
}
\caption{Abundance patterns of light neutron-capture elements from Sr to Pd of five observed stars and HD 122563 (black solid circle) compared with the predicted elemental abundances by the models of (a): and electron-capture supernova \citep{wana11} with various minimal electron fractions ($ Y_\mathrm{e, min}$) and of proto-neutron-star wind \citep{wana13} with various neutron-star masses ($M_\mathrm{NS}$) and minimal electron fractions of (b) $ Y_\mathrm{e, min}=0.40$ and (c) 0.45. The abundance patterns for the stars and the models are vertically shifted to match the Zr abundance of HD 122563.}

\label{ecsn}
\end{figure*}
%-----------------------------------------------------------------------------------------------------------------------------

As found in Figure \ref{lighte}, the abundance patterns of the five stars analyzed here and the two stars previously studied show a similar trend from Sr to Mo. The diversity becomes evident from Ru. As shown in Figure \ref{ZrMoRuPd} (a), the [Mo/Zr] ratios of all stars are within the range of random errors, while those of [Ru/Zr] show a wide distribution. This figure suggests that the abundance patterns are robust from Zr to Mo, while Ru does not show correlation with Mo and Zr. Figure \ref{ZrMoRuPd} (b) shows [Ru/Zr] abundance ratios as a function of [Pd/Zr]. A correlation is found in the five target stars and HD~122563 as indicated by the solid regression line. The slope of the regression line is less than one, indicating that abundance spreads are more wider in Pd than in Ru. Interestingly, the value of CS 22892-052 also locates near the extension of the regression line. In Figures \ref{ZrMoRuPd} (a) and (b), abundance ratios determined by \citet{hans14} for stars with [Fe/H]$<-2$ are also plotted by asterisks. Most of their results are in between the results of our target stars and that of CS 22892-052. Their results distribute along the slopes estimated for our sample, supporting the above argument about the diversity of the abundance pattern found for elements heavier than Mo.

\subsection{Comparison with Supernova Nucleosynthesis Models}
Based on the analysis of abundance patterns of our sample, we attempt to interpret the site of the weak r-process. In this subsection, we inspect the abundance patterns of our target stars by comparing them to nucleosynthesis models of types of supernovae, which are expected to be candidates of the weak r-process sites. Proposed candidates of the r-process site include supernovae \citep[e.g.,][]{woo94} and neutron-star mergers \citep[e.g.,][]{frei99}. Recent theoretical studies suggest traditional neutrino-driven supernovae as a major site of light r-process elements, but not for the heavy r-process elements \citep{wana11, wana13}. A scenario of magnetically driven supernovae has been proposed as a possible site of the main r-process, though nucleosynthetic outcomes are highly dependent on their poorly constrained physical conditions \citep{winteler12, nishimura15}. The latest nucleosynthesis studies based on general-relativistic hydrodynamical simulations show that neutron-star mergers can be a main r-process sites because they make both light and heavy r-process elements \citep{wana14, goriely15}. Additional support for the merger scenario has also been presented from the studies of spectroscopic analyses \citep[``r-process galaxy,"][]{ji16, roed16} and galactic chemical evolution \citep{tsujimoto14, hirai15, ishim15, shen15, vandevoort15}. For these reasons, we focus on the two available models of neutrino-driven supernovae \citep{wana11, wana13} to be compared with our result, though other scenarios may not be ruled out as weak r-process sites.

The supernova nucleosynthesis models by  \citet{wana11} and \citet{wana13} show diversity in elemental abundances by variations of parameters such as electron fraction and proto-neutron-star (PNS) mass. The model abundance patterns are scaled to Zr, as shown in colored lines of Figure~\ref{ecsn}, and most of the model lines show variations from the intermediate mass elements such as Ru. As shown in Section 4.3, we also find such diversity of the abundance patterns for light neutron-capture elements of metal-poor stars. While all the stars show similar trends from Sr to Mo, Ru abundances show variations; some stars show low Ru abundance compared to Mo, while some stars show higher abundance than Mo. We expect that this variation in Mo and Ru abundance ratios can serve as a hint to constrain the physical condition in supernovae. 

Figure~\ref{ecsn} (a) shows a comparison of our target stars with the two-dimensional hydrodynamical model (up to $\sim$ 350 ms after core bounce) of an electron-capture supernova \citep[ECSN,][]{wana11}. ECSNe are neutrino-driven explosions from collapsing oxygen-neon-magnesium cores of the low-mass end of supernova progenitors \citep{nomoto87, kitaura06, janka08, wana09}. The unique structure of a pre-supernova core with a relatively small barionic mass of $1.375\, M_\odot$ (gravitational mass of $\sim 1.2\, M_\odot$) surrounded by a dilute hydrogen-helium envelope leads to a fast explosion, ejecting less neutrino-processed, and thus, more neutron-rich material (than those expected for more massive progenitors). The original self-consistently exploding model of \citet{wana11}, however, results in the production of the lightest
neutron-capture elements up to Zr (red line in Figure~\ref{ecsn}(a)). They argued that a possibility of ejecting slightly more neutron-rich material could not be excluded because of the limitations of their simulation (e.g., resolution and two-dimensionality). As they demonstrated, the abundance pattern of light neutron-capture elements for HD 122563 can be reasonably explained when the original minimum electron fraction (proton-to-nucleon ratio) of $Y_\mathrm{e, min} = 0.40$ is replaced by $Y_\mathrm{e, min} = 0.30$ (blue line in Figure~\ref{ecsn} (a)). We, therefore, compare our result to the abundance patterns of their modified models with various $Y_\mathrm{e, min}$ values as shown in Figure~\ref{ecsn} (a), where the abundances are vertically shifted to match the Zr of HD 122563. Here, $Y_\mathrm{e, min} \gtrsim 0.3$ could be taken as a realistic range. Given the fairly robust physical conditions of ECSNe, the possible range of $Y_\mathrm{e, min}$ would be small in reality. Owing to the nucleosynthetic outcomes extremely sensitive to $Y_\mathrm{e}$, we find that the variations of abundance patterns in our samples can be well bracketed with a small range of $Y_\mathrm{e, min} = 0.25$--0.33. These models cannot explain, however, the abundance ratios Mo/Ru$~\gtrsim 1$ of the presented stars. Such a ratio is given by the model with $Y_\mathrm{e, min} = 0.35$; however, the absolute values of abundances are too small compared to the target stars.

Figure~\ref{ecsn} (b) and (c) show the comparison with the semi-analytic, spherically symmetric, general-relativistic model of PNS wind of core-collapse supernovae \citep{wana13}. The model involves a gravitational mass of proto-neutron star ($M_\mathrm{NS}$) and a $Y_\mathrm{e, min}$ as free parameters. The neutrino-driven wind phase lasts 10~s after core bounce with (somewhat arbitrary determined) temporal evolutions of neutrino luminosity, PNS radius, and $Y_\mathrm{e}$ \citep[see Figs.~1 and 2 in][]{wana13}. In Figures \ref{ecsn} (b) and (c), we compare our result with the abundance patterns obtained from the PNS models of $Y_\mathrm{e, min}= 0.40$ and 0.45, respectively. A reasonable range may be $Y_\mathrm{e, min} \gtrsim 0.4$ according to recent hydrodynamical studies \citep{fischer12, martinez12, roberts12}, which is, however, highly uncertain because of a lack of long-term multi-dimensional simulations with neutrino transport.

The colored lines show those of $M_\mathrm{NS} = 1.2$--$2.4\, M_\odot$. The abundances of the stars and of models are vertically shifted to match the Zr abundance of HD 122563. A realistic range of PNS masses would be $M_\mathrm{NS} \sim 1.2$--$1.6\, M_\odot$ according to recent systematic studies of one-dimensional core-collapse simulations \citep{ertl16}. As can be seen in Figure~\ref{ecsn} (b) ($Y_\mathrm{e, min} = 0.40$), the abundances of the stars are reasonably bracketed with a smaller range of $M_\mathrm{NS} = 1.2$--$1.4\, M_\odot$ \citep[than that in][]{ertl16}. If a greater $Y_\mathrm{e, min}$ (=0.45) is taken as in Figure~\ref{ecsn} (c), the range shifts to the more massive side of $M_\mathrm{NS} = 1.6$--$1.8\, M_\odot$, which is slightly out of that predicted from hydrodynamical studies \citep{ertl16}. Among the presented models, those with $Y_\mathrm{e, min} = 0.45$ and $M_\mathrm{NS} \le 1.6\, M_\odot$ satisfy the ratios Mo/Ru~$\lesssim 1$ of the stars. It suggests, therefore, that a reasonable parameter space of $M_\mathrm{NS}$ and $Y_\mathrm{e, min}$ brackets the abundance variations.

In conclusion, the diversity of abundance patterns in our sample can be interpreted as a result of possible ranges of quantities, such as $Y_\mathrm{e}$ and $M_\mathrm{NS}$, in relevant astrophysical sites. The variation of abundance patterns for the presented stars can be well explained with only small ranges of such quantities, e.g., $Y_\mathrm{e, min} = 0.25$--0.33 (ECSN models) or $M_\mathrm{NS} = 1.2$--$1.4\, M_\odot$ (a subset of PNS models). The result implies that the weak r-process site has fairly robust physical conditions (such as ECSNe), provided that our sample well represents the diversity of light neutron-capture abundances. Alternatively, the site has various physical conditions (such as PNS winds) and the diversity of neutron-capture abundances is in fact greater than that found in this study. The ratio Mo/Ru~$\lesssim 1$ can also serve as a constraint on theoretical modeling.

\section{Summary}
We analyzed the abundance patterns of five very-metal-poor stars, including Mo, Ru, and Pd, which are known to have high light-to-heavy abundance ratios of neutron-capture elements. We compared the abundances with the r-process component of the solar-system material and found that the patterns of light neutron-capture elements of our targets have diversity. We found a sizable variation in the abundances of elements starting from Ru toward Pd. We also confirmed robustness of abundance patterns up to Mo, which resulted in a small variation of abundance ratios of [Mo/Zr] as a function of [Ru/Zr] for the seven stars including our targets and reference (HD 122563 and CS 22892-052) stars. The significant decrease in light neutron-capture elements especially from Mo to Pd can be explained as a result of weak r-processing in core-collapse supernovae. In addition, the diversity found in our sample can be attributed to small variations of quantities, such as electron fraction or PNS mass, in relevant weak r-process sites. Future spectroscopic studies of additional weak r-process-like stars will be crucial to elucidate the diverseness of light neutron-capture abundances, which serve as diagnostics of the proposed supernova models.

\acknowledgments
The project was supported by the RIKEN iTHES Project, and JSPS
Grants-in-Aid for Scientific Research (26400232, 26400237). This work was also supported by JSPS and CNRS under the Japan-France Research Cooperative Program.

\end{document}